\documentclass[journal]{IEEEtran}
\ifdefined\pdfminorversion\pdfminorversion=7\fi
\IEEEoverridecommandlockouts
\usepackage{times,amsmath,amssymb,amsfonts,graphicx,cite,balance}
\usepackage{upgreek,bm}
\usepackage{algorithm,algorithmic}
\usepackage{booktabs,multirow}
\usepackage{hyperref}
\usepackage[caption=false,font=footnotesize]{subfig}
\DeclareMathOperator{\Tr}{Tr}
\DeclareMathSymbol{\upell}{\mathord}{ugrf@m}{"60}
\newcommand{\vect}[1]{\mathbf{#1}}

\newtheorem{assumption}{\upshape\textbf{Assumption}}

\newtheorem{theorem}{\upshape\textbf{Theorem}}

\newtheorem{proposition}{\upshape\textbf{Proposition}}

\begin{document}

\title{Joint Antenna Positioning and Dynamic Beam Tracking for MA-Enabled Full-Duplex Secure ISAC Systems: A Flow Matching DRL Approach}

\author{Zhendong Li, Yujie Zhao, Jianle Ba, Zhou Su, and Wen Chen
\thanks{Zhendong Li is with the School of Information and Communication Engineering, Xi'an Jiaotong University, Xi'an 710049, China (email: lizhendong@xjtu.edu.cn). Yujie Zhao is with the the School of Software, Xi’an Jiaotong University, Xi’an 710049, China (email: zyj@stu.xjtu.edu.cn). Jianle Ba and Zhou Su are with the School of Cyber Science and Engineering, Xi'an Jiaotong University, Xi'an 710049, China (email: bajianle2024@163.com; zhousu@ieee.org). Wen Chen is with the School of Integrated Circuits, Shanghai Jiao Tong University, Shanghai 200240, China (email: wenchen@sjtu.edu.cn). (Corresponding author: Zhou Su)}\vspace{-1.5em}
}

\maketitle

\begin{abstract}
Integrated sensing and communication (ISAC) improves spectral efficiency but faces significant security challenges in dynamic wireless environments, where the potential eavesdropper is treated as the sensing target. This paper proposes a movable antenna (MA)-enabled full-duplex secure ISAC system that tracks the mobile eavesdropper while safeguarding both uplink and downlink transmissions. Specifically, an extended Kalman filter (EKF) is employed to recursively estimate and predict the eavesdropper's state from radar echoes. Building on the estimated target state, we formulate a joint optimization problem to maximize the sum of the uplink and downlink secrecy rates by jointly optimizing the transmit and receive MA positions, uplink and downlink beamforming, artificial noise, and uplink transmit power. The resulting problem is challenging due to its inherent nonconvexity, strong coupling among optimization variables, and time-varying nature. Conventional iterative optimization methods may incur prohibitive computational overhead in dynamic environments, whereas existing deep reinforcement learning (DRL) methods based on unimodal Gaussian or deterministic policies may struggle to capture complex multimodal high-reward action distributions. To address these challenges, we develop a flow matching-based deep reinforcement learning (FMDRL) algorithm for sequential joint decision-making. The proposed algorithm integrates a conditional velocity-field actor with twin critics, generating continuous actions through finite-step Euler integration and employing a deterministic decoder to enforce physical constraints on antenna positions, transmit power, and positive semidefiniteness. Extensive numerical simulations confirm that the proposed FMDRL algorithm achieves superior performance in terms of both communication security and target tracking capabilities.
\end{abstract}


\section{Introduction}
\IEEEPARstart{W}{ith} the advancement of sixth-generation (6G) wireless networks, emerging applications such as intelligent transportation\cite{Liu2020VehicularISAC}, low-altitude networks\cite{Jiang2025LowAltitudeISAC}, and unmanned systems\cite{Meng2024UAVISAC} have imposed increasingly stringent requirements on high-speed reliable communications and high-accuracy environmental sensing. Consequently, integrated sensing and communication (ISAC) has attracted considerable attention\cite{Liu2022ISAC}. By jointly optimizing communication and sensing resources, ISAC systems enable communication and sensing functionalities to operate collaboratively and efficiently over the same spectrum resources, thereby achieving efficient spectrum utilization and enhanced system performance. However, ISAC systems also face security threats from open wireless channels and broadcast signals\cite{Su2021SecureISAC}. Eavesdroppers may intercept transmissions to obtain legitimate users' confidential information, causing data leakage and challenging system security.

Currently, extensive researches have been conducted on security issues in ISAC systems\cite{Su2022SecureInterference,Ren2023RobustSecureISAC,Bazzi2024SecureFD,Su2024SensingEstimation}.
The authors in\cite{Su2022SecureInterference} jointly designed transmit waveforms and receive beamformers to improve radar performance while suppressing target eavesdropping through interference exploitation.
In\cite{Su2024SensingEstimation}, sensing-assisted eavesdropper direction estimation was combined with artificial noise optimization to balance secrecy rate and estimation accuracy.
However, in dynamic wireless environments, rapidly changing channel conditions pose new challenges to secure beamforming design.
Security issues in ISAC systems in dynamic environments have also been investigated\cite{Liu2023MobileUplink,AlHabob2024PredictiveISAC,Zhang2025SecureISACformer,AlHabob2025PredictiveISAC,Xu2025MobileNearField,Hua2026DynamicFD}.
Specifically,\cite{Liu2023MobileUplink} employed an extended Kalman filter (EKF) and alternating optimization to track a mobile eavesdropper and improve uplink secrecy performance.
Authors developed a transformer-based predictive beamforming scheme to enhance secure transmission while tracking eavesdroppers\cite{Zhang2025SecureISACformer}.
Meanwhile,\cite{Hua2026DynamicFD} combined EKF-based mobile-user tracking with robust beamforming and power allocation to secure full-duplex ISAC transmissions.
Notably, the above studies are mainly based on fixed position antenna (FPA) arrays. Such arrays limit the system's ability to cope with dynamic environments and make it difficult to flexibly adapt to rapidly changing channel conditions. Therefore, how to further improve the security of ISAC systems in dynamic environments remains an urgent problem to be addressed.

Recently, movable antenna (MA) have been proposed as a new antenna paradigm \cite{Zhu2024MAOverview}, and are also referred to as fluid antenna (FA) under different antenna architecture implementation scenarios \cite{New2025FASTutorial}. Unlike FPA arrays, MA can change their coordinates in the physical environment through mechanical movement or other means, thereby reconfiguring the wireless channel and improving system performance \cite{Ma2024MIMOCapacity}. By introducing additional spatial degrees of freedom (DoFs), MA can provide new opportunities for enhancing the sensing and communication performance of ISAC systems \cite{Lyu2025MAISAC,Jiang2025MAISAC,Qin2024MACRB,Yang2026MultistaticSecureMAISAC,Li2026MAFullDuplexISAC,Chen2026TensorAFDMISAC}.
Specifically, \cite{Lyu2025MAISAC} jointly optimized antenna positions and beamforming coefficients to enhance communication and sensing performance, while \cite{Jiang2025MAISAC} investigated the joint optimization of transmit and receive MA positions and beamforming to improve sensing performance.
In addition, some studies have also shifted toward empowering secure ISAC systems with MA technology \cite{Li2026GMLSecureMAISAC,Chen2026SensingSecureMAISAC,Li2026SecureFAISAC,Salem2026SecureFASISAC}.
Authors developed a meta learning-based framework for antenna positioning and beamforming in MA secure ISAC systems~\cite{Li2026GMLSecureMAISAC}.
In \cite{Chen2026SensingSecureMAISAC}, sensing-assisted eavesdropper direction estimation was combined with robust beamforming and MA position optimization to maximize the worst-case secrecy rate under sensing uncertainty.
Authors employed block successive upper-bound minimization to jointly optimize antenna positions and beamforming for sum secrecy rate maximization under sensing constraints\cite{Li2026SecureFAISAC}.
Meanwhile, \cite{Salem2026SecureFASISAC} combined fractional programming and successive convex approximation for joint precoding and FA port selection to improve secrecy performance while satisfying radar sensing requirements.
Furthermore, regarding research on MA secure ISAC in dynamic scenarios, \cite{Hung2025DRLSecureMAISAC} employed deep reinforcement learning (DRL) to adapt transmit beamforming and MA positions to environmental variations and imperfect eavesdropper channel state information (CSI).
From a vehicle to everything perspective, authors discussed MA-supported antenna pattern regulation and sensing-communication coupling interference as promising means of enhancing physical layer security\cite{Yu2025SecureV2XMAISAC}.

Notably, dynamic secure ISAC optimization is challenging due to the strong coupling among optimization variables and the resulting non-convexity. Time-varying eavesdropper states and channel conditions further demand robust optimization. Although MA improve system performance, the coupling between MA positions and beamforming increases the problem dimensionality, while movement limits and minimum spacing constraints further complicate dynamic joint optimization. Conventional algorithms such as alternating optimization based on convex approximation typically incur high computational costs and initialization sensitivity, making real-time optimization challenging in dynamic environments. Therefore, in recent years, DRL has been widely applied to solve such complex optimization problems \cite{Hung2025DRLSecureMAISAC}. However, some existing continuous action DRL methods typically parameterize stochastic policies using Gaussian distributions, whereas the high reward action distributions in practical optimization problems may exhibit multimodal characteristics. A unimodal Gaussian policy can hardly characterize high reward action modes accurately, thereby limiting the expressiveness of the policy and the optimization performance. To overcome these limitations, generative artificial intelligence offers a promising approach to enhancing DRL policy expressiveness. In particular, flow matching (FM) \cite{Lipman2023FlowMatching} enables flexible modeling of complex multimodal action distributions, overcoming the limitations of  unimodal Gaussian policies \cite{Park2025FlowQLearning}. By integrating FM with DRL \cite{McAllister2026FlowMatchingPolicyGradients}, more expressive policies can be learned, providing a promising solution for joint MA positioning and dynamic tracking with a dynamic eavesdropper.

%
Motivated by above discussions, this paper focuses on the maximization problem of sum secrecy rate for  MA secure ISAC systems with a dynamic eavesdropper. The base station (BS) is assumed to operate in full-duplex mode and is equipped with two sets of MA for transmission and reception, respectively, where the MA is modeled as a continuous linear array. The objective is to maximize the sum secrecy rate by jointly optimizing the transmit and receive antenna positions, uplink and downlink beamforming, artificial noise, and uplink transmit power. Considering the non-convexity, time-varying nature, and strong coupling among variables, this paper proposes a flow matching deep reinforcement learning (FMDRL) optimization framework. Specifically, the framework establishes a mapping between actions and optimization variables. Moreover, a reward function is designed to incorporate the problem constraints and guide algorithm training. Unlike DRL methods that rely on Gaussian distributions, the proposed FMDRL method can better fit the multimodal high reward action distribution in the optimization problem, thereby enhancing the policy expressiveness and final optimization performance. The key contributions of this paper are outlined as follows:

\begin{itemize}
    \item We establish an MA-enabled full-duplex secure ISAC architecture in which the potential eavesdropper is treated as the sensing target. Both transmit and receive arrays support continuous linear MA positioning, and an EKF tracks the eavesdropper using radar echoes. We formulate a sum secrecy rate maximization problem by jointly optimizing MA positioning, uplink and downlink beamforming, artificial noise, and uplink transmit power under sensing, secrecy, and physical constraints. Solving this problem is challenging due to its inherent nonconvexity, the strong coupling among variables, and the temporal variations induced by eavesdropper mobility.

    \item To solve above challenges, we develop an FMDRL framework for joint antenna positioning and dynamic beam tracking. A conditional flow actor flexibly models multimodal action distributions through Euler integration, guided by twin critics and flow-matching regularization. Deterministic decoding enforces antenna geometry, uplink power bounds, and positive semidefiniteness of the artificial-noise covariance, while reward penalties address BS power, sensing, and secrecy requirements. The resulting policy coordinates spatial and transmission decisions to balance communication security and tracking performance in dynamic eavesdropping environments.
    
    \item Furthermore, we evaluate FMDRL through simulations covering communication performance, sensing performance, and sensitivity to flow steps. Regarding communication performance, the comparisons with other baselines demonstrate higher sum secrecy rate across different parameter settings. In terms of sensing performance, tracking errors and beampatterns are provided
    to confirm that the proposed FMDRL achieves the highest communication performance while simultaneously maintaining satisfactory sensing performance. Finally, the parameter experiments on FMDRL illustrate the flow from noise to the action distribution and reveal the impact of different algorithm parameters on the algorithm performance.

\end{itemize}

The remainder of this paper is organized as follows. Section~\ref{sec:system} presents the MA-enabled full-duplex secure ISAC system model, the EKF-based eavesdropper tracking scheme, and the formulation of the sum secrecy rate maximization problem. Building on this model, Section~\ref{sec:fmdrl} formulates the sequential decision problem and develops the FMDRL algorithm with constraint handling. Section~\ref{sec:simulation} then evaluates the proposed approach with baselines in terms of communication security, tracking accuracy, and solver sensitivity. Finally, Section~\ref{sec:conclusion}  concludes the paper.

\textit{Notations:} $a$, $\mathbf{a}$, and $\vect A$, denot a scalar, a vector, and a matric, respectively. $(\cdot)^{\mathrm T}$, $(\cdot)^{\mathrm H}$, and $(\cdot)^{-1}$ denote transpose, conjugate transpose, and matrix inverse, respectively. $\Tr(\vect A)$ denotes the trace of $\vect A$, and $\vect A\succeq\vect0$ means that $\vect A$ is Hermitian positive semidefinite. $[\vect A]_{i,j}$ and $[\vect a]_i$ denote the $(i,j)$-th entry of $\vect A$ and the $i$-th entry of $\vect a$, respectively. $\|\cdot\|_2$ denotes the Euclidean norm, while $|\cdot|$ denotes the absolute value of a scalar or the cardinality of a set, depending on the argument. $\operatorname{diag}(\vect a)$ constructs a diagonal matrix from $\vect a$, whereas $\operatorname{diag}(\vect A)$ extracts the diagonal entries of $\vect A$ as a vector. $\operatorname{vec}(\vect A)$ denotes vectorization, and $\operatorname{vec}_{\mathbb R}(\vect A)$ stacks the real and imaginary components into a real vector. $\mathbb E[\cdot]$ denotes expectation. $\mathbb R$ and $\mathbb C$ denote the sets of real and complex numbers, respectively. $\vect I_n$ denotes the $n\times n$ identity matrix, and $\vect0$ denotes a zero vector or matrix of appropriate dimensions. $[x]^+=\max\{x,0\}$, and $\operatorname{clip}(x,l,u)=\min\{\max\{x,l\},u\}$, applied componentwise to vector arguments.

\section{System Model and Problem Formulation}\label{sec:system}
We consider an MA-enabled full-duplex ISAC BS serving $U$ uplink users and $D$ downlink users in the presence of a dynamic eavesdropper, which is also the sensing target. The BS employs $N_{\rm rx}$ receive MA and $N_{\rm tx}$ transmit MA that share the same time and frequency resources. At slot $t$, the positions of the $m$-th receive MA and the $n$-th transmit MA are denoted by $\mathbf{p}_{\mathrm{rx},m}[t]=[x_{\mathrm{rx},m}[t],0]^{\mathrm{T}}$ and $\mathbf{p}_{\mathrm{tx},n}[t]=[x_{\mathrm{tx},n}[t],0]^{\mathrm{T}}$, respectively. Hence, both arrays can move continuously along the $x$-axis. We define $\mathbf{P}_{\rm rx}[t]$ and $\mathbf{P}_{\rm tx}[t]$ by stacking their corresponding MA locations.
\begin{figure}[t]
  \centering
  \includegraphics[width=\columnwidth]{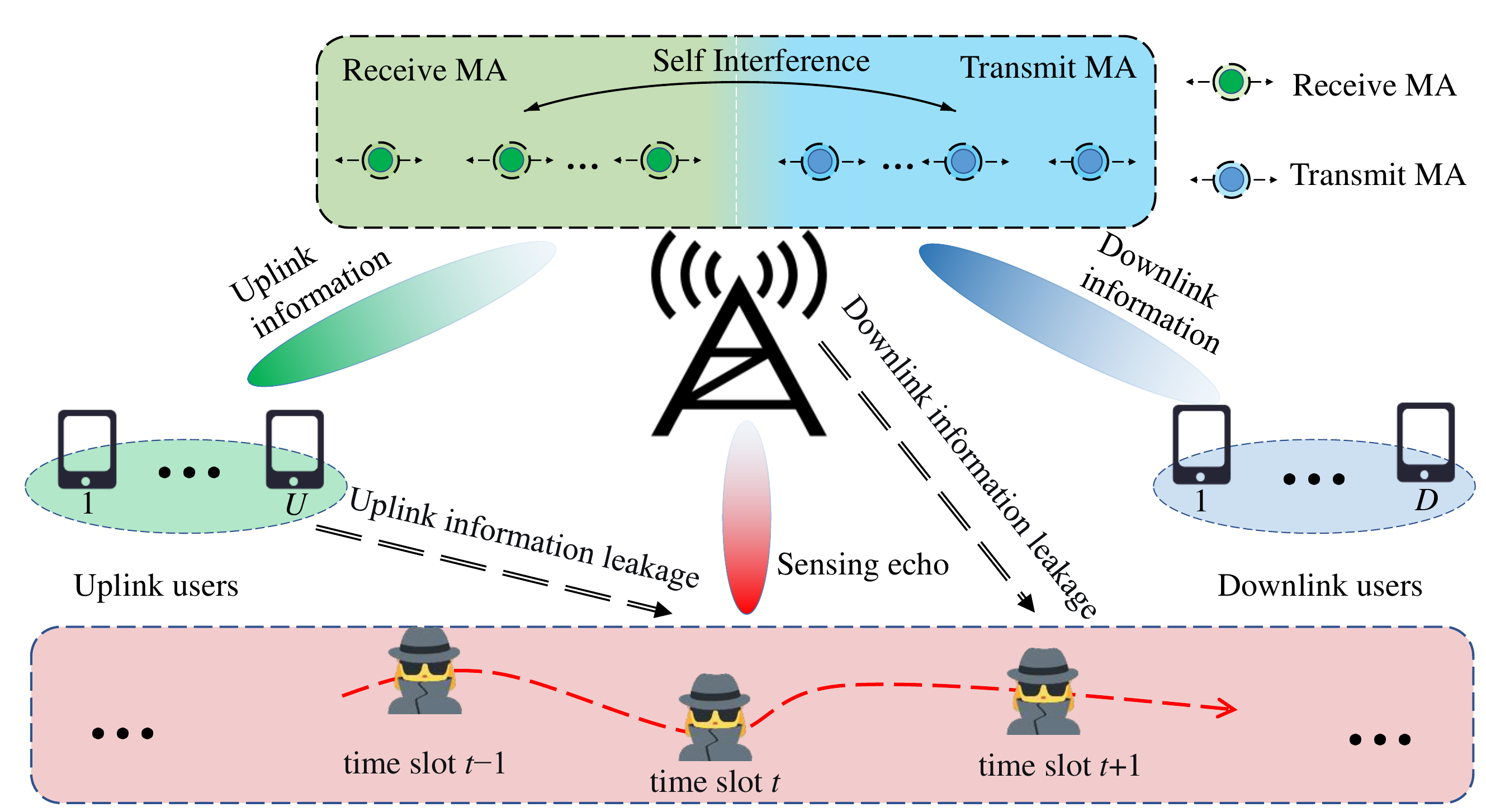}
  \caption{MA secure ISAC system with dynamic eavesdropper.}
  \label{fig:system_model}
  \vspace{-1.5em}
\end{figure}

The receive and transmit MA are respectively confined to the linear regions
\begin{equation}
\begin{aligned}
\mathcal C_{\rm rx}&=\{[x,0]^{\mathrm{T}}\mid 0\leq x\leq A_{\rm rx}\},\\
\mathcal C_{\rm tx}&=\{[x,0]^{\mathrm{T}}\mid 0\leq x\leq A_{\rm tx}\},
\end{aligned}
\end{equation}
where $A_{\rm rx}$ and $A_{\rm tx}$ are the available aperture lengths. To avoid mutual coupling, the distance between every two MA in the same array must be no less than $d_{\min}$. The uplink and downlink users are stationary, whereas the eavesdropper moves over time. All terminals are assumed to lie in the far field of the BS arrays. Let $\mathbf{u}_{{\rm u},k}$, $\mathbf{u}_{{\rm d},l}$, and $\mathbf{u}_{\rm e}[t]$ denote the locations of uplink user $k$, downlink user $l$, and the eavesdropper, respectively.

\subsection{Channel Model}
For a terminal at location $\mathbf{u}$, define the steering vector of an MA array with location matrix $\mathbf{P}$ as
\begin{equation}
\bm{\upalpha}(\mathbf{P},\mathbf{u})=
\left[e^{j\frac{2\pi}{\lambda}\mathbf{d}_{\rm dir}^{\mathrm{T}}\mathbf{p}_1},\ldots,
e^{j\frac{2\pi}{\lambda}\mathbf{d}_{\rm dir}^{\mathrm{T}}\mathbf{p}_N}\right]^{\mathrm{T}},
\quad \mathbf{d}_{\rm dir}=\frac{\mathbf{u}}{\|\mathbf{u}\|_2},
\label{eq:array_response}
\end{equation}
where $\lambda$ is the carrier wavelength. Accordingly, the downlink channel of user $l$ is modeled as
\begin{equation}
\begin{aligned}
\mathbf{h}_{{\rm d},l}[t]&=\sqrt{C_0\left(\frac{\|\mathbf{u}_{{\rm d},l}\|_2}{D_0}\right)^{-o}}
\bigg(\sqrt{\frac{\kappa_{{\rm d},l}}{1+\kappa_{{\rm d},l}}}\bm{\upalpha}(\mathbf{P}_{\rm tx}[t],\mathbf{u}_{{\rm d},l})\\
&\quad+\sqrt{\frac{1}{1+\kappa_{{\rm d},l}}}\mathbf{h}_{{\rm d},l}^{\rm NLoS}[t]\bigg),
\end{aligned}
\end{equation}
where $C_0$, $D_0$, $o$, and $\kappa_{{\rm d},l}$ are the reference path loss, reference distance, path-loss exponent, and Rician factor, respectively. $\mathbf{h}_{{\rm d},l}^{\rm NLoS}[t]$ has independent entries distributed as $\mathcal{CN}(0,1)$. The eavesdropper channel to the BS is
\begin{equation}
\begin{aligned}
\mathbf{h}_{\rm e}[t]
&=\sqrt{C_0\left(\frac{\|\mathbf{u}_{\rm e}[t]\|_2}{D_0}\right)^{-o}}
\bigg(
\sqrt{\frac{\kappa_{\rm e}}{1+\kappa_{\rm e}}}
\bm{\upalpha}(\mathbf{P}_{\rm tx}[t],\mathbf{u}_{\rm e}[t])
\\
&\qquad+\sqrt{\frac{1}{1+\kappa_{\rm e}}}
\mathbf{h}_{\rm e}^{\rm NLoS}[t]
\bigg),
\end{aligned}
\end{equation}
where $\kappa_{\rm e}$ is the Rician factor for the eavesdropper channel and $\mathbf{h}_{\rm e}^{\rm NLoS}[t] \sim \mathcal{CN}(0,1)$.
Similarly, the uplink channel from user $k$ is
\begin{equation}
\begin{aligned}
\mathbf{h}_{{\rm u},k}[t]&=\sqrt{C_0\left(\frac{\|\mathbf{u}_{{\rm u},k}\|_2}{D_0}\right)^{-o}}
\bigg(\sqrt{\frac{\kappa_{{\rm u},k}}{1+\kappa_{{\rm u},k}}}\bm{\upalpha}(\mathbf{P}_{\rm rx}[t],\mathbf{u}_{{\rm u},k})\\
&\quad+\sqrt{\frac{1}{1+\kappa_{{\rm u},k}}}\mathbf{h}_{{\rm u},k}^{\rm NLoS}[t]\bigg).
\end{aligned}
\end{equation}
The channel from uplink user $k$ to the eavesdropper is 
\begin{equation}
\begin{aligned}
g_{{\rm e},k}[t]
&=\sqrt{C_0\left(
\frac{\|\mathbf{u}_{{\rm u},k}-\mathbf{u}_{\rm e}[t]\|_2}{D_0}
\right)^{-o}}
\bigg(
\sqrt{\frac{\kappa_{\rm e}}{1+\kappa_{\rm e}}}
\\
&\qquad\cdot
e^{-j\frac{2\pi}{\lambda}
\|\mathbf{u}_{{\rm u},k}-\mathbf{u}_{\rm e}[t]\|_2}
+\sqrt{\frac{1}{1+\kappa_{\rm e}}}
\widetilde g_{{\rm e},k}^{\rm NLoS}[t]
\bigg),
\end{aligned}
\end{equation}
where $\widetilde g_{{\rm e},k}^{\rm NLoS}[t]\sim\mathcal{CN}(0,1)$. Due to full-duplex operation, the residual self-interference (SI) channel $\mathbf{H}_{\rm SI}[t]$ is considered, whose $(m,n)$-th entry is
\begin{equation}
[\mathbf{H}_{\rm SI}[t]]_{m,n}=\sqrt{\eta^{\rm SI}_{m,n}}
e^{-j2\pi\|\mathbf{p}_{{\rm rx},m}[t]-\mathbf{p}_{{\rm tx},n}[t]\|_2/\lambda}.
\end{equation}

\subsection{Signal Model}
At slot $t$, the BS transmits the ISAC signal
\begin{equation}
\mathbf{x}[t]=\sum_{l=1}^{D}\mathbf{w}_l[t]s_{{\rm d},l}[t]+\mathbf{z}[t],
\end{equation}
where $\mathbf{w}_l[t]\in\mathbb C^{N_{\rm tx}\times1}$ is the beamforming vector for downlink user $l$, $\mathbb E\{|s_{{\rm d},l}[t]|^2\}=1$, and $\mathbf{z}[t]\sim\mathcal{CN}(\bm{0},\mathbf{Z}[t])$ is artificial noise with $\mathbf{Z}[t]\succeq\mathbf{0}$. The received downlink signal is $y_{{\rm d},l}[t]=\mathbf{h}_{{\rm d},l}^{\mathrm{H}}[t]\mathbf{x}[t]+n_{{\rm d},l}[t]$, where $n_{{\rm d},l}[t]\sim\mathcal{CN}(0,\sigma_{{\rm d},l}^2)$.

Since the eavesdropper also acts as the sensing target, its echo received at the BS is
\begin{equation}
\begin{aligned}
\mathbf{y}_{\rm r}[t]&=\beta_{\rm r}[t]\bm{\upalpha}(\mathbf{P}_{\rm rx}[t],\mathbf{u}_{\rm e}[t])
\bm{\upalpha}^{\mathrm{H}}(\mathbf{P}_{\rm tx}[t],\mathbf{u}_{\rm e}[t])\\
&\quad\cdot\mathbf{x}[t-\tau[t]]e^{j2\pi\nu[t] t}+\mathbf{n}_{\rm r}[t],
\end{aligned}
\end{equation}
where $\beta_{\rm r}[t]$, $\tau[t]$, and $\nu[t]$ are the echo gain, round-trip delay, and Doppler shift, respectively, and $\mathbf{n}_{\rm r}[t]\sim\mathcal{CN}(\bm{0},\sigma_{\rm r}^2\mathbf{I})$. The BS receives uplink signals, SI, and the target echo concurrently
\begin{equation}
\mathbf{y}_{\rm b}[t]=\sum_{k=1}^{U}\sqrt{p_{{\rm u},k}[t]}\mathbf{h}_{{\rm u},k}[t]s_{{\rm u},k}[t]
+\mathbf{H}_{\rm SI}[t]\mathbf{x}[t]+\mathbf{y}_{\rm r}[t]+\mathbf{n}_{\rm b}[t],
\end{equation}
where $\mathbb E\{|s_{{\rm u},k}[t]|^2\}=1$ and $\mathbf{n}_{\rm b}[t]\sim\mathcal{CN}(\bm{0},\sigma_{\rm b}^2\mathbf{I})$. With receive beamforming $\mathbf{f}_k[t]$, the detected signal is $y_{{\rm u},k}[t]=\mathbf{f}_k^{\mathrm{H}}[t]\mathbf{y}_{\rm b}[t]$. Meanwhile, the eavesdropper receives
\begin{equation}
y_{\rm e}[t]=\mathbf{h}_{\rm e}^{\mathrm{H}}[t]\mathbf{x}[t]+\sum_{k=1}^{U}\sqrt{p_{{\rm u},k}[t]}g_{{\rm e},k}[t]s_{{\rm u},k}[t]+n_{\rm e}[t],
\end{equation}
where $n_{\rm e}[t]\sim\mathcal{CN}(0,\sigma_{\rm e}^2)$.

\subsection{EKF-Based Dynamic Beam Tracking}\label{sec:ekf}
In each slot, the eavesdropper’s state is tracked via monostatic radar echoes. After directional spatial filtering at the BS, matched filtering with the transmitted ISAC waveform yields the round-trip delay and Doppler shift, while a standard direction-of-arrival algorithm estimates the azimuth. Thus, the radar measurement vector is
\begin{equation}
\mathbf{y}_{\rm mea}[t]=[\hat\tau[t],\hat\nu[t],\hat\theta_{\rm mea}[t]]^{\mathrm{T}}
=\mathbf{h}_{\rm obs}(\bm{\upxi}_{\rm e}[t])+\mathbf{n}_{\rm mea}[t],
\label{eq:ekf_measurement}
\end{equation}
where the eavesdropper state is $\bm{\upxi}_{\rm e}[t]=[v_{\rm e}[t],\theta_{\rm e}[t],d_{\rm e}[t]]^{\mathrm{T}}$, containing radial speed, azimuth, and range. The nonlinear measurement function is
\begin{equation}
\mathbf{h}_{\rm obs}(\bm{\upxi}_{\rm e}[t])=
\begin{bmatrix}
2d_{\rm e}[t]/c\\
2f_cv_{\rm e}[t]\cos(\theta_{\rm e}[t])/c\\
\theta_{\rm e}[t]
\end{bmatrix},
\label{eq:ekf_observation_function}
\end{equation}
where $c$ and $f_c$ denote the speed of light and carrier frequency, respectively. The measurement noise is modeled as $\mathbf{n}_{\rm mea}[t]\sim\mathcal N(\bm{0},\mathbf{R}[t])$, with
\begin{equation}
\mathbf{R}[t]=\operatorname{diag}(\sigma_\tau^2[t],\sigma_\nu^2[t],\sigma_\theta^2[t]).
\label{eq:ekf_measurement_covariance}
\end{equation}
The delay and Doppler variances are inversely related to the matched-filtering echo signal-to-noise ratio (SNR), while $\sigma_\theta^2[t]$ captures the angular estimation error.

We adopt an approximately constant-velocity motion model over one slot interval $\Delta t$. Its nonlinear state transition is
\begin{equation}
\bm{\upxi}_{\rm e}[t+1]=\mathbf{f}_{\rm dyn}(\bm{\upxi}_{\rm e}[t])+\mathbf{q}[t],
\quad \mathbf{q}[t]\sim\mathcal N(\bm{0},\mathbf{Q}),
\label{eq:ekf_state_evolution}
\end{equation}
where $\mathbf{Q}=\operatorname{diag}(\sigma_v^2,\sigma_{\theta,q}^2,\sigma_{d,q}^2)$ accounts for motion-model mismatch and
\begin{equation}
\mathbf{f}_{\rm dyn}(\bm{\upxi}_{\rm e}[t])=
\begin{bmatrix}
v_{\rm e}[t]\\
\theta_{\rm e}[t]+\dfrac{v_{\rm e}[t]\Delta t\sin\theta_{\rm e}[t]}{d_{\rm e}[t]}\\
d_{\rm e}[t]-v_{\rm e}[t]\Delta t\cos\theta_{\rm e}[t]
\end{bmatrix}.
\label{eq:ekf_state_transition}
\end{equation}

Let $\hat{\bm{\upxi}}_{\rm e}[t|t]$ and $\mathbf{C}[t|t]$ be the posterior state estimate and its error covariance after processing slot $t$ measurements. The EKF prediction is
\begin{align}
\hat{\bm{\upxi}}_{\rm e}[t|t-1]&=\mathbf{f}_{\rm dyn}(\hat{\bm{\upxi}}_{\rm e}[t-1|t-1]),\\
\mathbf{C}[t|t-1]&=\mathbf{J}_{\rm dyn}[t-1]\mathbf{C}[t-1|t-1]\mathbf{J}_{\rm dyn}^{\mathrm{T}}[t-1]+\mathbf{Q},
\label{eq:ekf_prediction}
\end{align}
where $\mathbf{J}_{\rm dyn}[t-1]$ is the state-transition Jacobian. For $\hat v=\hat v_{\rm e}[t-1|t-1]$, $\hat\theta=\hat\theta_{\rm e}[t-1|t-1]$, and $\hat d=\hat d_{\rm e}[t-1|t-1]$, it is given by
\begin{equation}
\mathbf{J}_{\rm dyn}[t-1]=\begin{bmatrix}
1&0&0\\
\dfrac{\Delta t\sin\hat\theta}{\hat d}&1+\dfrac{\hat v\Delta t\cos\hat\theta}{\hat d}&-\dfrac{\hat v\Delta t\sin\hat\theta}{\hat d^2}\\
-\Delta t\cos\hat\theta&\hat v\Delta t\sin\hat\theta&1
\end{bmatrix}.
\label{eq:ekf_transition_jacobian}
\end{equation}
The observation Jacobian, evaluated at $\hat{\bm{\upxi}}_{\rm e}[t|t-1]$, is
\begin{equation}
\mathbf{J}_{\rm obs}[t]=\begin{bmatrix}
0&0&2/c\\
\dfrac{2f_c}{c}\cos\bar\theta&-\dfrac{2f_c}{c}\bar v\sin\bar\theta&0\\
0&1&0
\end{bmatrix},
\label{eq:ekf_observation_jacobian}
\end{equation}
where $\bar v=\hat v_{\rm e}[t|t-1]$ and $\bar\theta=\hat\theta_{\rm e}[t|t-1]$. The innovation covariance and Kalman gain are
\begin{align}
\mathbf{S}[t]&=\mathbf{J}_{\rm obs}[t]\mathbf{C}[t|t-1]\mathbf{J}_{\rm obs}^{\mathrm{T}}[t]+\mathbf{R}[t],\\
\mathbf{K}[t]&=\mathbf{C}[t|t-1]\mathbf{J}_{\rm obs}^{\mathrm{T}}[t]\big(\mathbf{S}[t]\big)^{-1}.
\label{eq:ekf_gain}
\end{align}
Thus, the measurement update becomes
\begin{align}
\hat{\bm{\upxi}}_{\rm e}[t|t]&=\hat{\bm{\upxi}}_{\rm e}[t|t-1]+\mathbf{K}[t]\big(\mathbf{y}_{\rm mea}[t]-\mathbf{h}_{\rm obs}(\hat{\bm{\upxi}}_{\rm e}[t|t-1])\big),\\
\mathbf{C}[t|t]&=(\mathbf{I}_3-\mathbf{K}[t]\mathbf{J}_{\rm obs}[t])\mathbf{C}[t|t-1].
\label{eq:ekf_update}
\end{align}
The updated state is then propagated once more to obtain $\hat{\bm{\upxi}}_{\rm e}[t+1|t]$ and $\mathbf{C}[t+1|t]$, which are used for the next-slot antenna positioning and dynamic beam tracking design. We quantify tracking quality through the predicted mean-squared error $\Tr(\mathbf{C}[t+1|t])$.

\subsection{Secure Communication Metrics}
The downlink user signal-to-interference-plus-noise ratio (SINR) and its eavesdropping counterpart are
\begin{align}
\gamma_{{\rm d},l}[t]&=\frac{|\mathbf{h}_{{\rm d},l}^{\mathrm{H}}[t]\mathbf{w}_l[t]|^2}{\sum_{i\ne l}|\mathbf{h}_{{\rm d},l}^{\mathrm{H}}[t]\mathbf{w}_i[t]|^2+\mathbf{h}_{{\rm d},l}^{\mathrm{H}}[t]\mathbf{Z}[t]\mathbf{h}_{{\rm d},l}[t]+\sigma_{{\rm d},l}^2},\\
\gamma_{{\rm e},l}^{\rm DL}[t]&=\frac{|\mathbf{h}_{\rm e}^{\mathrm{H}}[t]\mathbf{w}_l[t]|^2}{\sum_{i\ne l}|\mathbf{h}_{\rm e}^{\mathrm{H}}[t]\mathbf{w}_i[t]|^2+\mathbf{h}_{\rm e}^{\mathrm{H}}[t]\mathbf{Z}[t]\mathbf{h}_{\rm e}[t]+\sigma_{\rm e}^2}.
\end{align}
The corresponding downlink secrecy rate is $R_{{\rm s},l}^{\rm DL}[t]=[\log_2(1+\gamma_{{\rm d},l}[t])-\log_2(1+\gamma_{{\rm e},l}^{\rm DL}[t])]^+$. Let $P_{\rm tx}[t]=\sum_{l=1}^{D}\|\mathbf{w}_l[t]\|_2^2+\Tr(\mathbf{Z}[t])$. The uplink SINRs at the BS and the eavesdropper are given by
\begin{align}
\gamma_{{\rm u},k}[t]&=
\frac{p_{{\rm u},k}[t]|\mathbf{f}_k^{\mathrm{H}}[t]\mathbf{h}_{{\rm u},k}[t]|^2}
{\begin{aligned}
\sum_{i\ne k}p_{{\rm u},i}[t]|\mathbf{f}_k^{\mathrm{H}}[t]\mathbf{h}_{{\rm u},i}[t]|^2
&+\|\mathbf{f}_k^{\mathrm{H}}[t]\mathbf{H}_{\rm SI}[t]\|_2^2P_{\rm tx}[t]\\
&+\sigma_{\rm b}^2\|\mathbf{f}_k[t]\|_2^2
\end{aligned}},\\
\gamma_{{\rm e},k}^{\rm UL}[t]&=\frac{p_{{\rm u},k}[t]|g_{{\rm e},k}[t]|^2}{\sum_{i\ne k}p_{{\rm u},i}[t]|g_{{\rm e},i}[t]|^2+\mathbf{h}_{\rm e}^{\mathrm{H}}[t]\mathbf{Z}[t]\mathbf{h}_{\rm e}[t]+\sigma_{\rm e}^2}.
\end{align}
Therefore, $R_{{\rm s},k}^{\rm UL}[t]=[\log_2(1+\gamma_{{\rm u},k}[t])-\log_2(1+\gamma_{{\rm e},k}^{\rm UL}[t])]^+$.

\subsection{Problem Formulation}
At each slot, the BS first predicts the eavesdropper state using the EKF and then jointly optimizes transmit and receive MA positions, uplink and downlink beamforming, artificial noise, and uplink transmit power. Let $\mathbf{W}[t]=[\mathbf{w}_1[t],\ldots,\mathbf{w}_D[t]]$, $\mathbf{F}[t]=[\mathbf{f}_1[t],\ldots,\mathbf{f}_U[t]]$, and $\mathbf{p}_{\rm u}[t]=[p_{{\rm u},1}[t],\ldots,p_{{\rm u},U}[t]]^{\mathrm{T}}$. The per-slot optimization problem is
\begin{subequations}\label{prob:P1}
\begin{align}
\text{(P1)}\quad\max_{\substack{
  \mathbf{W}[t], \mathbf{Z}[t], \mathbf{F}[t], \\
  \mathbf{p}_{\rm u}[t], \mathbf{P}_{\rm rx}[t], \mathbf{P}_{\rm tx}[t]
}}\nonumber
&\ \sum_{l=1}^{D}R_{{\rm s},l}^{\rm DL}[t]+\sum_{k=1}^{U}R_{{\rm s},k}^{\rm UL}[t]\\
\text{s.t.}\quad \Tr(\mathbf{C}[t+1|t])&\leq\varepsilon_{\max},\\
P_{\text{tx}}[t]&\leq P_{\rm BS}^{\max},\\
0\leq p_{{\rm u},k}[t]&\leq P_{{\rm u},k}^{\max},\ \forall k,\\
R_{{\rm s},l}^{\rm DL}[t]&\geq\bar R_{{\rm d},l},\ \forall l,\\
R_{{\rm s},k}^{\rm UL}[t]&\geq\bar R_{{\rm u},k},\ \forall k,\\
\mathbf{p}_{{\rm rx},m}[t]\in\mathcal C_{\rm rx},\quad \mathbf{p}_{{\rm tx},n}[t]&\in\mathcal C_{\rm tx},\ \forall m,n,\\
\|\mathbf{p}_{{\rm rx},m_1}[t]-\mathbf{p}_{{\rm rx},m_2}[t]\|_2&\geq d_{\min},\ \forall m_1\ne m_2,\\
\|\mathbf{p}_{{\rm tx},n_1}[t]-\mathbf{p}_{{\rm tx},n_2}[t]\|_2&\geq d_{\min},\ \forall n_1\ne n_2,\\
\mathbf{Z}[t]&\succeq\mathbf{0}.
\end{align}
\end{subequations}
Here, $\varepsilon_{\max}$ is the sensing prediction-error threshold, $P_{\rm BS}^{\max}$ and $P_{{\rm u},k}^{\max}$ are the BS and uplink user power budgets, and $\bar R_{{\rm d},l}$ and $\bar R_{{\rm u},k}$ are the minimum downlink and uplink secrecy rate requirements, respectively. 
Solving problem~\eqref{prob:P1} directly is intractable as it is time-varying, non-convex, and involves tightly coupled variables.
Specifically, the secrecy rate objective is generally nonconcave due to the logarithmic SINR differences and the coupling among beamforming, artificial noise, and transmit power. Moreover, the nonlinear dependence of the channels on antenna positions and the nonconvex minimum spacing constraints further complicate the joint optimization.

\section{FMDRL-Based Joint Antenna Positioning and Dynamic Beam Tracking Algorithm}
\label{sec:fmdrl}
\begin{figure}[t]
  \centering
  \includegraphics[width=\columnwidth]{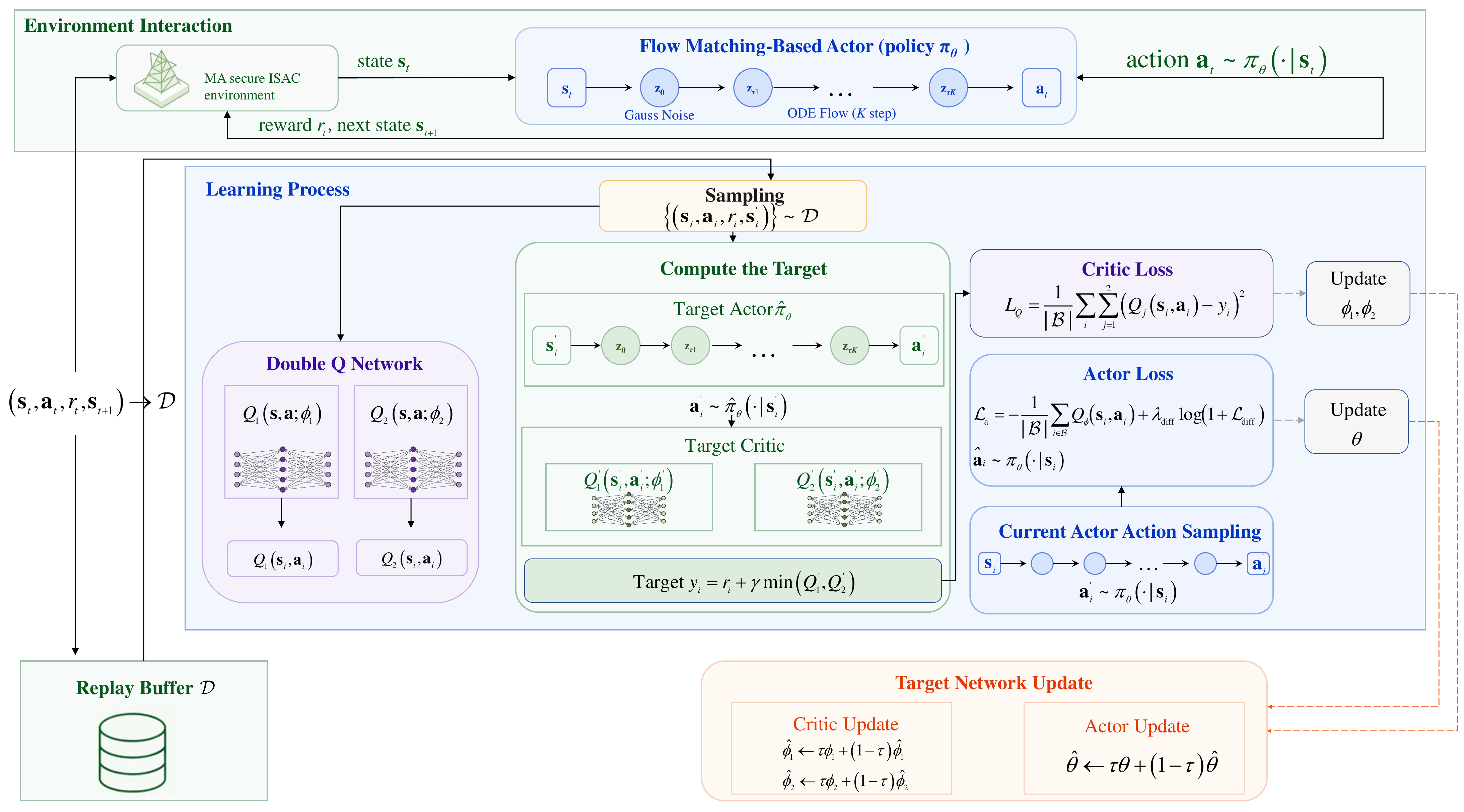}
  \caption{Overall flowchart of the proposed FMDRL algorithm.}
  \label{fig:fmdrl_flow_chart}
   \vspace{-1.5em}
\end{figure}

To address above challenges, we develop a FMDRL framework. We first formulate a sequential decision model and then construct an online off-policy twin-Q actor-critic algorithm in which a conditional velocity field actor generates bounded actions via explicit Euler integration and replay action velocity matching. The complete flowchart is depicted in Fig.~\ref{fig:fmdrl_flow_chart}.

\subsection{Sequential Decision Formulation}
We describe the decision process by $(\mathcal S,\mathcal A,\mathcal P,r,\gamma_{\rm RL})$, where $\mathcal S$, $\mathcal A$, $\mathcal P$, $r$, and $\gamma_{\rm RL}\in[\text{0},\text{1}]$ denote the state space, action space, transition kernel, reward function, and discount factor, respectively. Each decision corresponds to one physical time slot $t$. The reinforcement learning state is denoted by $\mathbf{s}[t]$.

\subsubsection{State Space}
Under partial observability, the state $\mathbf{s}[t]$ serves as a compact approximate information state comprising the channel features, prior antenna positions $\mathbf{P}_{\rm tx}^{-}[t], \mathbf{P}_{\rm rx}^{-}[t]$, and EKF prior covariance diagonal $\bm{\uppsi}[t]=\operatorname{diag}(\mathbf{C}[t|t-1])$. Let $\mathbf{H}_{\rm d}[t]=[\mathbf{h}_{{\rm d},1}[t],\ldots,\mathbf{h}_{{\rm d},D}[t]]^{\mathrm{H}}$ and $\mathbf{H}_{\rm u}[t]=[\mathbf{h}_{{\rm u},1}[t],\ldots,\mathbf{h}_{{\rm u},U}[t]]^{\mathrm{H}}$. Using the predicted eavesdropper position $\hat{\mathbf{u}}_{\rm e}[t]$ from $\hat{\bm{\upxi}}_{\rm e}[t|t-1]$, we construct the estimated eavesdropping channels $\widehat{\mathbf{h}}_{\rm e}[t]$, $\widehat{\mathbf{g}}_{\rm e}[t]=[g_{{\rm e},1}[t],\ldots,g_{{\rm e},U}[t]]^{\mathrm{T}}$, and the array responses $\bm{\upalpha}_{i}[t]\triangleq\bm{\upalpha}(\mathbf{P}_{i}^{-}[t],\hat{\mathbf{u}}_{\rm e}[t])$ for $i\in\{\mathrm{tx},\mathrm{rx}\}$. Stacking real and imaginary parts via $\operatorname{vec}_{\mathbb R}(\cdot)$, the raw channel feature vector is
\begin{equation}
\begin{aligned}
\mathbf{c}[t]=\big[&\operatorname{vec}_{\mathbb R}(\mathbf{H}_{\rm d}[t])^{\mathrm{T}},
\operatorname{vec}_{\mathbb R}(\mathbf{H}_{\rm u}[t])^{\mathrm{T}},
\operatorname{vec}_{\mathbb R}(\widehat{\mathbf{h}}_{\rm e}[t])^{\mathrm{T}},\\
&\operatorname{vec}_{\mathbb R}(\widehat{\mathbf{g}}_{\rm e}[t])^{\mathrm{T}},
\operatorname{vec}_{\mathbb R}\big(\bm{\upalpha}_{\rm tx}[t]\bm{\upalpha}_{\rm rx}[t]^{\mathrm{T}}\big)^{\mathrm{T}}\big]^{\mathrm{T}}.
\end{aligned}
\end{equation}
Applying component-wise normalization $\operatorname{scale}(\cdot)$ to each feature group, the state vector is defined as
\begin{equation}
\mathbf{s}[t]=\operatorname{scale}\Big(\big[\mathbf{c}[t]^{\mathrm{T}}, \operatorname{vec}(\mathbf{P}_{\rm tx}^{-}[t])^{\mathrm{T}}, \operatorname{vec}(\mathbf{P}_{\rm rx}^{-}[t])^{\mathrm{T}}, \bm{\uppsi}[t]^{\mathrm{T}}\big]^{\mathrm{T}}\Big).
\label{eq:fmdrl_state}
\end{equation}

\subsubsection{Action Space}
The bounded action vector $\mathbf{a}[t]\in\mathcal A$ comprises six blocks:
\begin{equation}
\mathbf{a}[t]=\big[\mathbf{a}_{\rm tx}^{\mathrm{T}}[t],\mathbf{a}_{\rm rx}^{\mathrm{T}}[t],
\mathbf{a}_{\rm w}^{\mathrm{T}}[t],\mathbf{a}_{\rm an}^{\mathrm{T}}[t],\mathbf{a}_{\rm f}^{\mathrm{T}}[t],
\mathbf{a}_{\rm p}^{\mathrm{T}}[t]\big]^{\mathrm{T}},
\label{eq:fmdrl_action}
\end{equation}
where $\mathbf{a}_{\rm tx}[t]\in[\text{0},\text{1}]^{N_{\rm tx}+1}$ and $\mathbf{a}_{\rm rx}[t]\in[\text{0},\text{1}]^{N_{\rm rx}+1}$ denote the transmit and receive antenna gap weights. $\mathbf{a}_{\rm w}[t]\in[-\text{1},\text{1}]^{2DN_{\rm tx}}$, $\mathbf{a}_{\rm an}[t]\in[-\text{1},\text{1}]^{2N_{\rm tx}}$, and $\mathbf{a}_{\rm f}[t]\in[-\text{1},\text{1}]^{2UN_{\rm rx}}$ represent the real and imaginary parts of the downlink beamforming vector, artificial-noise factor, and uplink beamforming vector, respectively. $\mathbf{a}_{\rm p}[t]\in[-\text{1},\text{1}]^U$ denotes the normalized uplink powers. A deterministic decoder maps $\mathbf{a}[t]$ to the feasible physical variables in~\eqref{prob:P1} as follows.

For an $N$-element array within $[x_{\min},x_{\min}+A]$, let $\bm{\upomega}=[\omega_0,\ldots,\omega_N]^{\mathrm{T}}\in[\text{0},\text{1}]^{N+1}$ denote the gap weight vector ($\mathbf{a}_{\rm tx}[t]$ or $\mathbf{a}_{\rm rx}[t]$), normalized into proportions $\varpi_i=\omega_i/\sum_{j=0}^{N}\omega_j$. Allocating the residual movable length $L_{\rm free}=A-(N-1)d_{\min}$ across boundary and inter-element gaps, the $n$-th antenna coordinate $[x_n,0]^{\mathrm{T}}$ is given by
\begin{equation}
x_n=x_{\min}+(n-1)d_{\min}+L_{\rm free}\sum_{i=0}^{n-1}\varpi_i,\quad n=1,\ldots,N,
\label{eq:fmdrl_position_mapping}
\end{equation}
which inherently guarantees aperture containment, minimum spacing $d_{\min}$, and element ordering.

Reconstructing complex entries from $\mathbf{a}_{\rm w}[t]$, $\mathbf{a}_{\rm an}[t]$, and $\mathbf{a}_{\rm f}[t]$ yields the unscaled matrices $\widetilde{\mathbf{W}}[t]$, $\widetilde{\bm{\upell}}[t]\in\mathbb C^{N_{\rm tx}\times1}$, and $\widetilde{\mathbf{F}}[t]$. With $P_{\rm pre}[t]=\|\widetilde{\mathbf{W}}[t]\|_{\rm F}^2+\|\widetilde{\bm{\upell}}[t]\|_2^2$ and scaling factor $\chi[t]=\min\big\{1,\sqrt{P_{\rm BS}^{\max}/P_{\rm pre}[t]}\big\}$, the BS transmit variables are obtained via
\begin{equation}
\mathbf{W}[t]=\chi[t]\widetilde{\mathbf{W}}[t],\quad
\bm{\upell}[t]=\chi[t]\widetilde{\bm{\upell}}[t],\quad
\mathbf{Z}[t]=\bm{\upell}[t]\bm{\upell}^{\mathrm{H}}[t],
\label{eq:fmdrl_an_mapping}
\end{equation}
satisfying $\mathbf{Z}[t]\succeq\mathbf{0}$ and the power budget $P_{\rm tx}[t]\leq P_{\rm BS}^{\max}$. Finally, each column of $\widetilde{\mathbf{F}}[t]$ is normalized to unit norm to obtain $\mathbf{f}_k[t]$, and the uplink powers are mapped via
\begin{equation}
p_{{\rm u},k}[t]=\frac{P_{{\rm u},k}^{\max}}{2}\big([\mathbf{a}_{\rm p}[t]]_k+1\big),\quad k=1,\ldots,U.
\label{eq:fmdrl_power_mapping}
\end{equation}

\subsubsection{State Transition}
After decoding the action, the environment updates the communication and SI channels. The resulting transmit signal and array geometry determine the sensing quality and hence the measurement uncertainty used in the EKF update. The posterior estimate is propagated to obtain the next slot prediction and $\mathbf{C}[t+1|t]$. Together with the updated channel and position features, these quantities form $\mathbf{s}[t+1]$. The transition is represented by
\begin{equation}
\mathbf{s}[t+1]\sim\mathcal P(\cdot\mid\mathbf{s}[t],\mathbf{a}[t]).
\end{equation}
In particular, the current action influences both the immediate secrecy performance and the information available for subsequent decisions.

\subsubsection{Reward Function}
We separate the strict secrecy rates $R_{{\rm s},l}^{\rm DL}[t]=[\Delta R_{{\rm d},l}[t]]^+$ and $R_{{\rm s},k}^{\rm UL}[t]=[\Delta R_{{\rm u},k}[t]]^+$  from the shaped training reward, where $\Delta R_{{\rm d},l}[t]=R_{{\rm d},l}[t]-R_{{\rm e},l}^{\rm DL}[t]$ and $\Delta R_{{\rm u},k}[t]=R_{{\rm u},k}[t]-R_{{\rm e},k}^{\rm UL}[t]$ denote the downlink and uplink rate differences. With $\ell_{\beta_{\rm leak}}(x)=x$ for $x\geq0$ and $\ell_{\beta_{\rm leak}}(x)=\beta_{\rm leak} x$ for $x<0$, define the communication reward and normalized constraint penalties as
\begin{equation}
\begin{aligned}
r_{\rm comm}[t]={}&\frac{1}{D}\sum_{l=1}^{D}\ell_{\beta_{\rm leak}}\big(\Delta R_{{\rm d},l}[t]\big)\\
&+\frac{1}{U}\sum_{k=1}^{U}\ell_{\beta_{\rm leak}}\big(\Delta R_{{\rm u},k}[t]\big),\\
\delta_{\rm c}[t]={}&\frac{[\Tr(\mathbf{C}[t+1|t])-\varepsilon_{\max}]^+}{\varepsilon_{\max}},\\
\delta_{\rm d}[t]={}&\frac{\big[\bar R_{\rm d,l}-\min_{l}R_{{\rm s},l}^{\rm DL}[t]\big]^+}{\bar R_{\rm d}},\\
\delta_{\rm u}[t]={}&\frac{\big[\bar R_{\rm u,k}-\min_{k}R_{{\rm s},k}^{\rm UL}[t]\big]^+}{\bar R_{\rm u}}.
\end{aligned}
\end{equation}
The instantaneous reward is
\begin{equation}
r[t]=r_{\rm comm}[t]-\eta_{\rm c}\delta_{\rm c}[t]
-\eta_{\rm d}\delta_{\rm d}[t]-\eta_{\rm u}\delta_{\rm u}[t],
\label{eq:fmdrl_reward}
\end{equation}
where $\eta_{\rm c}$, $\eta_{\rm d}$, and $\eta_{\rm u}$ are the penalty coefficients for the dynamic beam tracking constraint, downlink secrecy rate constraint, and uplink secrecy rate constraint, respectively.

\subsubsection{Learning Objective}
Over a $T$-slot horizon, the agent seeks an optimal policy $\pi$ that maximizes the expected discounted return
\begin{equation}
\max_{\pi}\ J(\pi)=\mathbb E_{\pi}
\left[\sum_{t=1}^{T}\gamma_{\rm RL}^{t-1}r[t]\right].
\label{eq:fmdrl_objective}
\end{equation}
This formulation recasts the intractable Problem~\eqref{prob:P1} into a sequential decision making process, balancing long term secrecy rate maximization with EKF tracking accuracy.

\subsection{FMDRL Optimization Framework}
\begin{figure}[t]
  \centering
  \includegraphics[width=\columnwidth]{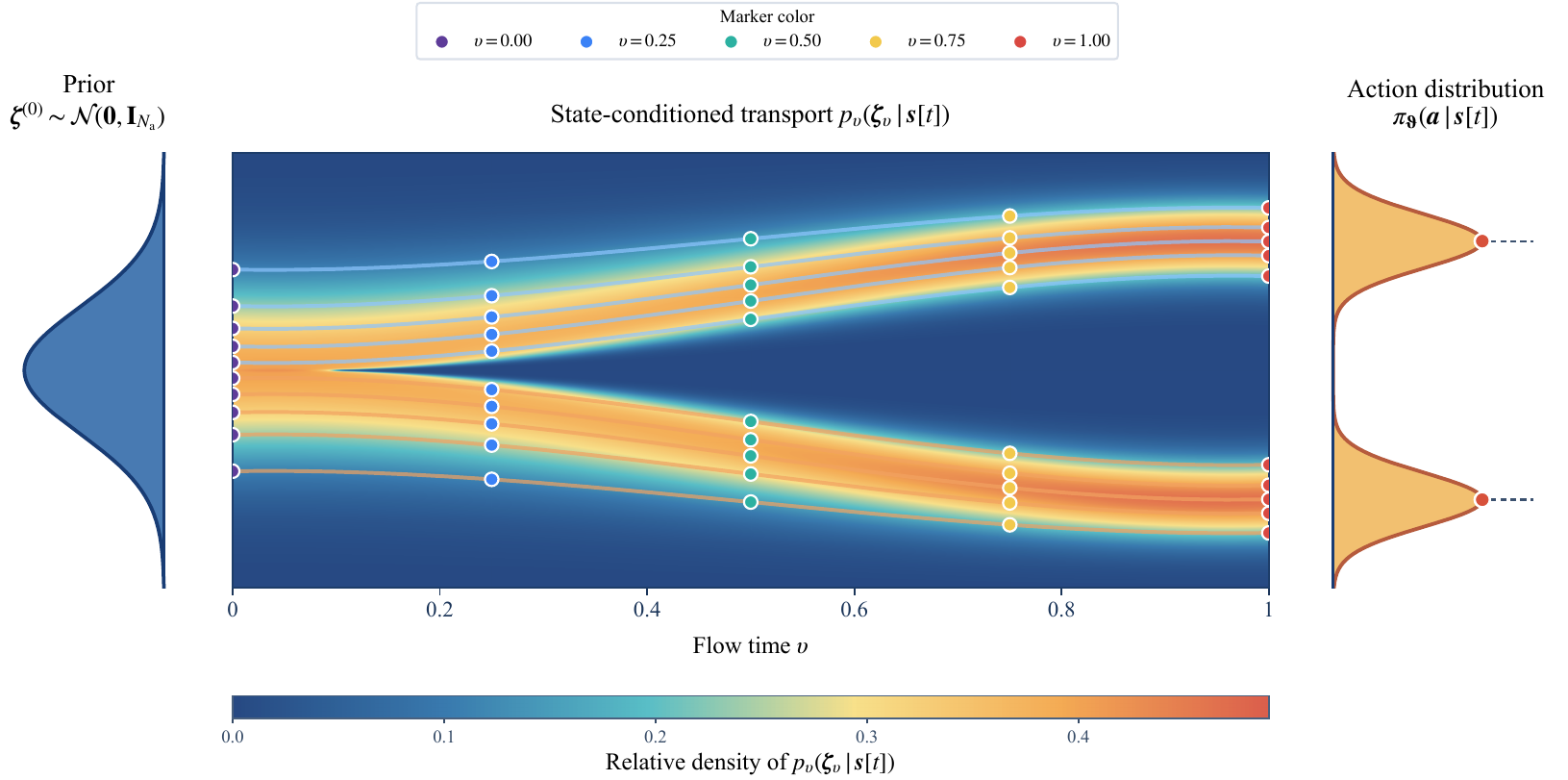}
  \caption{Conditional flow matching actor visualization.}
  \label{fig:flow_matching_principle}
  \vspace{-1.5em}
\end{figure}
The proposed framework combines a conditional velocity field network parameterized by $\bm{\upvartheta}$ with twin critics parameterized by $\bm{\upphi}=(\bm{\upphi}_1,\bm{\upphi}_2)$ and target parameters $\bar{\bm{\upvartheta}}$ and $\bar{\bm{\upphi}}=(\bar{\bm{\upphi}}_1,\bar{\bm{\upphi}}_2)$. The actor defines a stochastic policy $\pi_{\bm{\upvartheta}}(\cdot\mid\mathbf{s})$ for online data collection and a deterministic action map $\bm{\upmu}_{\bm{\upvartheta}}(\mathbf{s})$ for Q-optimization, target evaluation, and deployment.

\subsubsection{Conditional Flow Matching Actor}
Due to the non-convex coupling between antenna positioning and beamforming in~\eqref{prob:P1}, the optimal policy often exhibits a complex, multimodal action landscape that conventional unimodal Gaussian actors struggle to capture. To overcome this limitation, as illustrated in Fig.~\ref{fig:flow_matching_principle}, we construct a conditional flow matching actor that continuously transports a standard Gaussian prior $\mathcal N(\bm{0},\mathbf{I}_{N_{\rm a}})$ at flow time $\upsilon=\text{0}$ to a expressive, multimodal target action distribution at $\upsilon=\text{1}$ via a state-conditioned probability path. Specifically, the actor learns a velocity field $\mathbf{v}_{\bm{\upvartheta}}(\bm{\upzeta}_{\upsilon},\upsilon,\mathbf{s})$ over $\upsilon\in[\text{0},\text{1}]$ and flow variable $\bm{\upzeta}_{\upsilon}\in\mathbb R^{N_{\rm a}}$ by combining a state encoder, a sinusoidal time embedding, and an action fusion network.

Given replay actions $\mathbf{a}^{\rm buf}\in\mathcal A$ from buffer $\mathcal D$, we form unified flow endpoints $\bm{\upzeta}_1^{\rm buf}\in[-\text{1},\text{1}]^{N_{\rm a}}$ by affinely rescaling the position weight blocks as $\bm{\upzeta}_{1,{\rm pos}}^{\rm buf}=2\mathbf{a}_{\rm pos}^{\rm buf}-\bm{1}$ while keeping other blocks unchanged. Sampling $(\mathbf{s},\mathbf{a}^{\rm buf})\sim\mathcal D$, $\bm{\upepsilon}\sim\mathcal N(\bm{0},\mathbf{I}_{N_{\rm a}})$, and $\upsilon\sim\mathcal U(\text{0},\text{1})$, the linear interpolation path and target velocity are defined as
\begin{equation}
\bm{\upzeta}_{\upsilon}=(1-\upsilon)\bm{\upepsilon}+\upsilon\bm{\upzeta}_1^{\rm buf},\quad
\mathbf{v}_{\rm tar}=\bm{\upzeta}_1^{\rm buf}-\bm{\upepsilon}.
\label{eq:fmdrl_path}
\end{equation}
The velocity matching loss is given by
\begin{equation}
L_{\rm FM}(\bm{\upvartheta})=
\mathbb E_{\substack{(\mathbf{s},\mathbf{a}^{\rm buf})\sim\mathcal D\\
\upsilon,\bm{\upepsilon}}}
\left[\frac{1}{N_{\rm a}}\|\mathbf{v}_{\bm{\upvartheta}}(\bm{\upzeta}_{\upsilon},\upsilon,\mathbf{s})
-\mathbf{v}_{\rm tar}\|_2^2\right].
\label{eq:fmdrl_fm_loss}
\end{equation}

As illustrated in \textbf{Algorithm~\ref{alg:cfm_action}}, to generate an action in state $\mathbf{s}$, the actor integrates the ordinary differential equation (ODE) $\mathrm d\bm{\upzeta}_{\upsilon}/\mathrm d\upsilon=\mathbf{v}_{\bm{\upvartheta}}(\bm{\upzeta}_{\upsilon},\upsilon,\mathbf{s})$ from $\bm{\upzeta}^{(0)}$ over $K_{\rm FM}$ explicit Euler steps with step size $\Delta\upsilon=1/K_{\rm FM}$ and $\upsilon_j=j\Delta\upsilon$
\begin{equation}
\bm{\upzeta}^{(j+1)}=\bm{\upzeta}^{(j)}+\Delta\upsilon\,
\mathbf{v}_{\bm{\upvartheta}}(\bm{\upzeta}^{(j)},\upsilon_j,\mathbf{s}),
\quad j=0,\ldots,K_{\rm FM}-1.
\label{eq:fmdrl_euler}
\end{equation}
The terminal state $\bm{\upzeta}^{(K_{\rm FM})}$ is then mapped to the bounded action $\mathbf{a}=\mathcal T_{\rm out}(\bm{\upzeta}^{(K_{\rm FM})})\in\mathcal A$ via
\begin{equation}
[\mathbf{a}]_i=\begin{cases}
\big(\operatorname{clip}([\bm{\upzeta}^{(K_{\rm FM})}]_i,-1,1)+1\big)/2, & i\in\mathcal I_{\rm pos},\\[2pt]
\operatorname{clip}([\bm{\upzeta}^{(K_{\rm FM})}]_i,-1,1), & i\notin\mathcal I_{\rm pos},
\end{cases}
\label{eq:fmdrl_output}
\end{equation}
where $\mathcal I_{\rm pos}$ indexes the position weight entries.

\begin{algorithm}[t]
\caption{Conditional Flow Matching Action Generation}
\label{alg:cfm_action}
\begin{algorithmic}[1]
\REQUIRE State $\mathbf{s}$, velocity network $\mathbf{v}_{\bm{\upvartheta}}$, and Euler steps $K_{\rm FM}$.
\vspace{-1.em}
\STATE Sample prior $\bm{\upzeta}^{(0)}\sim\mathcal N(\bm{0},\mathbf{I}_{N_{\rm a}})$ and set $\Delta\upsilon=1/K_{\rm FM}$.
\FOR{$j=0,\ldots,K_{\rm FM}-1$}
    \STATE $\bm{\upzeta}^{(j+1)}\leftarrow\bm{\upzeta}^{(j)}+\Delta\upsilon\,\mathbf{v}_{\bm{\upvartheta}}(\bm{\upzeta}^{(j)},j\Delta\upsilon,\mathbf{s})$ via~\eqref{eq:fmdrl_euler}.
\ENDFOR
\STATE Obtain bounded action $\mathbf{a}=\mathcal T_{\rm out}(\bm{\upzeta}^{(K_{\rm FM})})$ via~\eqref{eq:fmdrl_output}.
\STATE Decode $\mathbf{a}$ into $\big\{\mathbf{P}_{\rm tx},\mathbf{P}_{\rm rx},\mathbf{W},\mathbf{Z},\mathbf{F},\mathbf{p}_{\rm u}\big\}$ via~\eqref{eq:fmdrl_position_mapping}--\eqref{eq:fmdrl_power_mapping}.
\ENSURE Action $\mathbf{a}\in\mathcal A$ and physical variables.
\end{algorithmic}
\end{algorithm}
\subsubsection{Twin-Critic Evaluation and Joint Optimization}
To evaluate actions with explicit geometric awareness, the twin critics share a state encoder and employ two value heads $\widetilde Q_{\bm{\upphi}_i}(\mathbf{s},\mathbf{a})\triangleq Q_{\bm{\upphi}_i}\big(\mathbf{s},\mathcal M_{\rm c}(\mathbf{a})\big)$, $i\in\{\text{1},\text{2}\}$. 

For an $h$-step trajectory segment with non-terminal bootstrap indicator $b_{\rm boot}[t,h]$, the Bellman target is
\begin{equation}
\begin{aligned}
y_{\rm TD}[t]={}&\sum_{j=0}^{h-1}\gamma_{\rm RL}^j r[t+j]+\gamma_{\rm RL}^h b_{\rm boot}[t,h]\\
&\cdot\min_{i\in\{1,2\}}\widetilde Q_{\bar{\bm{\upphi}}_i}
\big(\mathbf{s}[t+h],\bm{\upmu}_{\bar{\bm{\upvartheta}}}(\mathbf{s}[t+h])\big).
\end{aligned}
\label{eq:fmdrl_target}
\end{equation}
Given a mini-batch $\mathcal B\subset\mathcal D$, the critics are updated by minimizing the temporal-difference error:
\begin{equation}
L_{\rm c}(\bm{\upphi})=
\frac{1}{|\mathcal B|}\sum_{t\in\mathcal B}\sum_{i=1}^{2}
\big(\widetilde Q_{\bm{\upphi}_i}(\mathbf{s}[t],\mathbf{a}[t])-y_{\rm TD}[t]\big)^2.
\label{eq:fmdrl_critic_loss}
\end{equation}
With the critics fixed, the actor is trained end-to-end through $\mathcal M_{\rm c}$, $\mathcal T_{\rm out}$, and the $K_{\rm FM}$ Euler steps by minimizing the joint objective:
\begin{equation}
\begin{aligned}
L_{\rm actor}(\bm{\upvartheta})={}&-\frac{1}{|\mathcal B|}
\sum_{\mathbf{s}\in\mathcal B}\min_{i\in\{1,2\}}
\widetilde Q_{\bm{\upphi}_i}\big(\mathbf{s},\bm{\upmu}_{\bm{\upvartheta}}(\mathbf{s})\big)\\
&+\lambda_{\rm FM}\log\big(1+\widehat{L}_{\rm FM}(\bm{\upvartheta})\big),
\end{aligned}
\label{eq:fmdrl_actor_loss}
\end{equation}
where $\widehat{L}_{\rm FM}(\bm{\upvartheta})$ is the empirical estimate of~\eqref{eq:fmdrl_fm_loss} over $\mathcal B$ and $\lambda_{\rm FM}\geq0$ balances Q-value maximization with flow regularization.

\subsubsection{Training and Deployment}
Following random exploration steps, online transitions are continuously stored in $\mathcal D$, and both actor and critic networks are optimized via AdamW. Every $N_{\rm upd}$ critic updates, the actor is updated via~\eqref{eq:fmdrl_actor_loss}, followed by target network soft updates with $\rho\in(0,1)$
\begin{equation}
\bar{\bm{\upvartheta}}\leftarrow\rho\bm{\upvartheta}+(1-\rho)\bar{\bm{\upvartheta}},\quad
\bar{\bm{\upphi}}_i\leftarrow\rho\bm{\upphi}_i+(1-\rho)\bar{\bm{\upphi}}_i,\ \ i\in\{1,2\}.
\label{eq:fmdrl_soft_update}
\end{equation}
The complete training workflow is summarized in \textbf{Algorithm~\ref{alg:fmdrl_training}}. During training, the twin critics guide policy improvement, while flow matching regularization encourages the actor to capture the state-conditioned action patterns represented in the replay buffer.

\begin{algorithm}[H]
\caption{FMDRL Training Framework}
\label{alg:fmdrl_training}
\begin{algorithmic}[1]
\REQUIRE Horizon $T$, update interval $N_{\rm upd}$, and $\gamma_{\rm RL},\rho,\lambda_{\rm FM}$.
\vspace{-1em}
\STATE Initialize $\mathbf{v}_{\bm{\upvartheta}}$, $Q_{\bm{\upphi}_1},Q_{\bm{\upphi}_2}$, targets $\bar{\bm{\upvartheta}}\leftarrow\bm{\upvartheta},\bar{\bm{\upphi}}_i\leftarrow\bm{\upphi}_i$, and buffer $\mathcal D$.
\FOR{each episode}
    \STATE Reset environment and obtain $\mathbf{s}[1]$ via~\eqref{eq:fmdrl_state}.
    \FOR{$t=1,\ldots,T$}
        \STATE Generate $\mathbf{a}[t]$ and physical variables via \textbf{Algorithm~\ref{alg:cfm_action}}.
        \vspace{-1em}
        \STATE Execute action, update EKF, and observe $r[t]$, $\mathbf{s}[t+1]$, $b_{\rm boot}[t,h]$.
        \STATE Store transition in $\mathcal D$ and sample mini-batch $\mathcal B=\{(\mathbf{s},\mathbf{a}^{\rm buf},r,\mathbf{s}',b_{\rm boot})\}\subset\mathcal D$.
        \STATE Compute Bellman target $y_{\rm TD}$ via~\eqref{eq:fmdrl_target}.
        \STATE Update critics $\bm{\upphi}_1,\bm{\upphi}_2$ by minimizing $L_{\rm c}(\bm{\upphi})$ in~\eqref{eq:fmdrl_critic_loss}.
        \vspace{-1em}
        \IF{$t \bmod N_{\rm upd} = 0$}
            \STATE Form $\bm{\upzeta}_1^{\rm buf}$ from $\mathbf{a}^{\rm buf}$; sample $\upsilon\sim\mathcal U(0,1)$ and $\bm{\upepsilon}\sim\mathcal N(\bm{0},\mathbf{I}_{N_{\rm a}})$.
            \STATE Compute $\bm{\upzeta}_{\upsilon},\mathbf{v}_{\rm tar}$ via~\eqref{eq:fmdrl_path} and $\widehat{L}_{\rm FM}(\bm{\upvartheta})$ via~\eqref{eq:fmdrl_fm_loss}.
            \STATE Update actor $\bm{\upvartheta}$ by minimizing $L_{\rm actor}(\bm{\upvartheta})$ in~\eqref{eq:fmdrl_actor_loss}.
            \STATE Update targets $\bar{\bm{\upvartheta}},\bar{\bm{\upphi}}_1,\bar{\bm{\upphi}}_2$ via~\eqref{eq:fmdrl_soft_update}.
        \ENDIF
    \ENDFOR
\ENDFOR
\ENSURE Trained actor $\mathbf{v}_{\bm{\upvartheta}}$ and critics $Q_{\bm{\upphi}_1},Q_{\bm{\upphi}_2}$.
\end{algorithmic}
\end{algorithm}
\vspace{-1.em}
\subsubsection{Theoretical Analysis}
\label{sec:theory}
This section establishes the end-to-end gradient stability through the $K_{\rm FM}$-step Euler integration and the asymptotic convergence of the FMDRL framework.

\begin{assumption}
\label{assump:compact}
(i) The velocity field $\mathbf{v}_{\bm{\upvartheta}}(\bm{\upzeta},\upsilon,\mathbf{s})$ is $L_v$-Lipschitz in $\bm{\upzeta}$ with bounded parameter Jacobian $\|\nabla_{\bm{\upvartheta}}\mathbf{v}_{\bm{\upvartheta}}\|_{\rm op}\le G_{\vartheta}$. The composite output critic mapping $\mathcal M_{\rm c}\circ\mathcal T_{\rm out}$ and critics $Q_{\bm{\upphi}_i}$ are $L_T$- and $L_Q$-Lipschitz, respectively. 
(ii) The actor parameters $\bm{\upvartheta}_n$ are projected onto a compact convex set $\Theta$, and the  learning rates $\{\beta_n,\alpha_n\}$ satisfy standard two-timescale stochastic approximation conditions: $\sum_n\alpha_n=\sum_n\beta_n=\infty$, $\sum_n(\alpha_n^2+\beta_n^2)<\infty$, and $\alpha_n/\beta_n\to 0$, with target soft-update rate $\rho_n=c\beta_n$ ($c>0$).
\end{assumption}

\begin{proposition}[Step-Independent Gradient Stability]
\label{prop:grad_bound}
Under \textbf{Assumption~\ref{assump:compact}} (i), the actor gradients from both the Q-value pathway $\mathcal L_Q(\bm{\upvartheta})=-\mathbb E_{\mathbf{s}}\big[\min_{i\in\{1,2\}}\widetilde Q_{\bm{\upphi}_i}(\mathbf{s},\mathbf{a})\big]$ and the logarithmic flow regularization $\Psi_{\rm FM}(\bm{\upvartheta})=\log\big(1+L_{\rm FM}(\bm{\upvartheta})\big)$ are uniformly bounded as
\begin{equation}
\|\nabla_{\bm{\upvartheta}}\mathcal L_Q(\bm{\upvartheta})\|\le L_Q L_T G_{\vartheta} e^{L_v},\quad
\|\nabla_{\bm{\upvartheta}}\Psi_{\rm FM}(\bm{\upvartheta})\|\le \frac{G_{\vartheta}}{\sqrt{N_{\rm a}}}.
\label{eq:theory_gradient_bounds}
\end{equation}
Crucially, both bounds are independent of the Euler steps $K_{\rm FM}$.
\end{proposition}
\begin{IEEEproof}
Differentiating the Euler recursion~\eqref{eq:fmdrl_euler} with respect to $\bm{\upzeta}^{(j)}$ yields the step Jacobian $\mathbf{J}_j=\mathbf{I}_{N_{\rm a}}+K_{\rm FM}^{-1}\nabla_{\bm{\upzeta}}\mathbf{v}_{\bm{\upvartheta}}(\bm{\upzeta}^{(j)},\upsilon_j,\mathbf{s})$, which satisfies $\|\mathbf{J}_j\|_{\rm op}\le 1+L_v/K_{\rm FM}$. For any step $j$, the cumulative transition matrix $\mathbf{P}_j=\prod_{m=j+1}^{K_{\rm FM}-1}\mathbf{J}_m$ satisfies $\|\mathbf{P}_j\|_{\rm op}\le(1+L_v/K_{\rm FM})^{K_{\rm FM}-1-j}\le e^{L_v}$. Applying the chain rule through $\mathcal M_{\rm c}\circ\mathcal T_{\rm out}$ and averaging over $K_{\rm FM}$ steps yields the first bound in~\eqref{eq:theory_gradient_bounds}. Moreover, by Cauchy--Schwarz, $\|\nabla_{\bm{\upvartheta}}L_{\rm FM}(\bm{\upvartheta})\|\le\frac{2G_{\vartheta}}{\sqrt{N_{\rm a}}}\sqrt{L_{\rm FM}(\bm{\upvartheta})}$. Using the inequality $\frac{2\sqrt{u}}{1+u}\le 1$ for $u=L_{\rm FM}(\bm{\upvartheta})\ge 0$, the logarithmic barrier $\Psi_{\rm FM}(\bm{\upvartheta})$ strictly caps the flow gradient norm by $G_{\vartheta}/\sqrt{N_{\rm a}}$.
\end{IEEEproof}

\begin{theorem}[Two-Timescale Convergence]
\label{thm:convergence}
Under \textbf{Assumption~\ref{assump:compact}}, let $\bm{\Upupsilon}_n=(\bm{\upphi}_{1,n},\bm{\upphi}_{2,n},\bar{\bm{\upphi}}_{1,n},\bar{\bm{\upphi}}_{2,n},\bar{\bm{\upvartheta}}_n)$ denote the fast critic-target variables. Suppose that for any fixed $\bm{\upvartheta}\in\Theta$, the fast mean ODE admits a unique globally asymptotically stable equilibrium $\bm{\Upupsilon}^*(\bm{\upvartheta})$ that is Lipschitz in $\bm{\upvartheta}$, with critic component $\bm{\upphi}^*(\bm{\upvartheta})$. Define the slow projected actor drift and its stationary set as
\begin{equation}
\begin{aligned}
\mathcal H(\bm{\upvartheta})
&=\Pi_{T_\Theta(\bm{\upvartheta})}
\big[-\nabla_{\bm{\upvartheta}}L_{\rm actor}(\bm{\upvartheta})
\big|_{\bm{\upphi}=\bm{\upphi}^*(\bm{\upvartheta})}\big],\\
\mathcal S^*
&=\{\bm{\upvartheta}\in\Theta:
\mathcal H(\bm{\upvartheta})=\bm{0}\},
\end{aligned}
\label{eq:theory_slow_field}
\end{equation}
where $\Pi_{T_\Theta(\bm{\upvartheta})}[\cdot]$ denotes Euclidean projection onto the tangent cone of $\Theta$. Assume that the projected ODE $\dot{\bm{\upvartheta}}=\mathcal H(\bm{\upvartheta})$ is well posed and admits a continuously differentiable strict Lyapunov function $V$ for $\mathcal S^*$, with $V(\mathcal S^*)$ having empty interior. Then, almost surely as $n\to\infty$,
\begin{equation}
\|\bm{\Upupsilon}_n-\bm{\Upupsilon}^*(\bm{\upvartheta}_n)\|\longrightarrow 0,\qquad
\operatorname{dist}(\bm{\upvartheta}_n,\mathcal S^*)\longrightarrow 0.
\label{eq:theory_convergence}
\end{equation}
\end{theorem}
\begin{IEEEproof}
By \textbf{Proposition~\ref{prop:grad_bound}}, the stochastic actor gradients have bounded second moments, ensuring that the mini-batch gradient noise forms a martingale difference sequence with bounded variance. Since $\alpha_n/\beta_n\to 0$, the actor parameter $\bm{\upvartheta}_n$ acts as a quasi-static parameter on the fast timescale, driving $\bm{\Upupsilon}_n\to\bm{\Upupsilon}^*(\bm{\upvartheta}_n)$ almost surely via Borkar's two-timescale stochastic approximation theorem. Substituting the converged critic equilibrium $\bm{\upphi}^*(\bm{\upvartheta}_n)$ into the slow timescale shows that $\bm{\upvartheta}_n$ asymptotically tracks the projected ODE $\dot{\bm{\upvartheta}}=\mathcal H(\bm{\upvartheta})$, converging almost surely to the stationary set $\mathcal S^*$.
\end{IEEEproof}

\section{Numerical Results}\label{sec:simulation}
In this section, the numerical simulation results of the MA secure ISAC system with a dynamic eavesdropper are presented. The simulation parameters are set in Table~\ref{tab:table2}.
\begin{table}[t] 
    \centering 
    \caption{Simulation Parameters}
    \label{tab:table2} 
    \begin{tabular}{lc}
        \toprule
        Parameters  & Values   \\
        \midrule
        \multicolumn{2}{l}{\textbf{\textit{System Parameters}}} \\
        Carrier wavelength $\lambda$  & 0.1 $\text{m}$ \\
         Minimum distance between any two MA elements $d_\text{min}$ & 0.5 $\lambda$  \\
        Path loss exponent   $o$ & 2.2  \\
        Self-interference gain $\eta_{m,n}^\text{SI}$ & $-$100 dB\\
        Speed of the eavesdropper $v_\text{e}$ & 20 $\text{m/s}$  \\
        Total time slots $T$ & 50  \\
        Noise power  $\sigma^\text{2}$ & 1$\times\text{10}^\text{-7} \text{mW}$  \\
        \addlinespace 
        \multicolumn{2}{l}{\textbf{\textit{Algorithm Parameters}}} \\
        Number of training epochs $N_{\mathrm{e}}$ & 700 \\
        Actor learning rate $\alpha_{\mathrm{a}}$ & $\text{3}\times\text{10}^{-\text{4}}$ \\
        Critic learning rate $\alpha_{\mathrm{c}}$ & $\text{3}\times\text{10}^{-\text{4}}$ \\
        Soft update coefficient $\rho$ & $\text{5}\times\text{10}^{-\text{3}}$ \\
        Discount factor $\gamma_{\rm RL}$ & 0.99 \\
        Number of FM solver steps $K_{\rm FM}$ & 12 \\
        Batch size $B$ & 256 \\
        Downlink secrecy-rate penalty coefficient $\eta_{\rm d}$ & 2.0 \\
        Uplink secrecy-rate penalty coefficient $\eta_{\rm u}$ & 2.2 \\
        Sensing-error penalty coefficient $\eta_{\rm c}$ & 1.5 \\
        FM regularization coefficient $\lambda_{\rm FM}$ & 0.01 \\
        \bottomrule
    \end{tabular}
\end{table}

 Considering the characteristics of ISAC systems, this paper evaluates the performance from two aspects: communication performance and sensing performance, and compares it with other baseline algorithms. We also examine the flow matching action generation process and analyze the sensitivity of system performance to the number of solver steps. The baseline algorithms are enumerated as follows:
\begin{itemize}
    \item \textbf{Baseline 1} (SAC): Soft actor-critic (SAC)\cite{Haarnoja2018SAC} is a maximum-entropy reinforcement learning algorithm with a stochastic policy.
    \item \textbf{Baseline 2} (DDPG): Deep deterministic policy gradient (DDPG)\cite{Lillicrap2016DDPG} is an actor-critic reinforcement learning algorithm with a deterministic policy.
    \item \textbf{Baseline 3} (TD3): Twin delayed deep deterministic policy gradient (TD3)\cite{Fujimoto2018TD3} extends DDPG with twin critics, delayed policy updates, and target policy smoothing.
    \item \textbf{Baseline 4} (FPA): SAC is applied with fixed transmit and receive antenna arrays.
\end{itemize}
\subsection{Communication Performance}
In this subsection, the proposed algorithm FMDRL is compared with several baseline algorithms in terms of communication metrics. All numerical simulation results are obtained over $S$ random seeds. 

\begin{figure}[H]
  \centering
  \includegraphics[width=0.9\columnwidth]{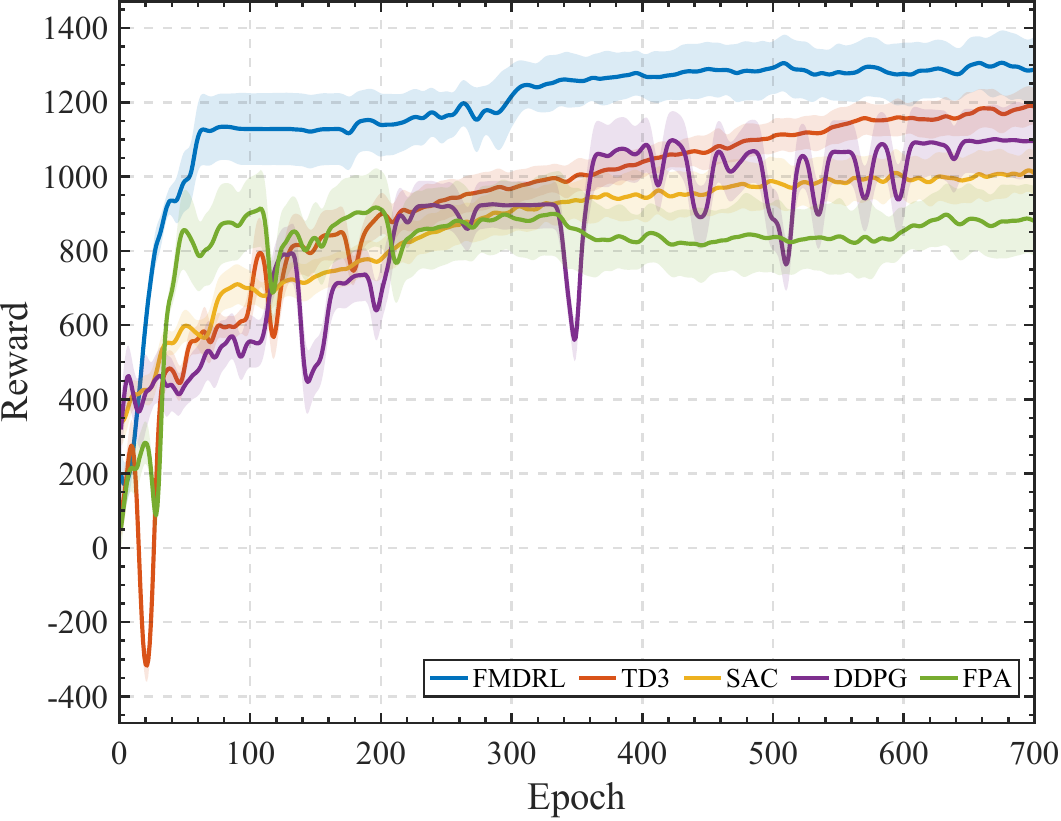}
  \caption{Reward versus training epochs across different algorithms.}
  \label{fig:reward_vs_epoch}
  \vspace{-1em}
\end{figure}

We first present the training reward comparisons of all algorithms. Fig.~\ref{fig:reward_vs_epoch} illustrates the reward versus the training epochs over $\text{700}$ epochs, where the solid curves and shaded regions represent the mean rewards and standard deviations across random seeds, respectively. As observed in Fig.~\ref{fig:reward_vs_epoch}, the cumulative rewards of all algorithms generally increase as training progresses and eventually reach stationary plateaus. It is evident that FPA converges to the lowest reward plateau of approximately $\text{879.4}$ over the final $\text{50}$ epochs due to its spatially static array geometry, which deprives the BS of antenna positioning DoFs for dynamic channel reconfiguration. In contrast, the four MA algorithms achieve substantially higher rewards by dynamically adjusting the transmit and receive MA positions, demonstrating the pronounced spatial diversity enabled by antenna mobility. Among the four MA optimization algorithms, the proposed FMDRL exhibits  the highest performance, climbing rapidly within the first $\text{80}$ epochs and converging to an average reward of $\text{1297.3}$, which is $\text{9.93}\%$, $\text{18.30}\%$, $\text{28.97}\%$, and $\text{47.52}\%$ higher than those of TD3, DDPG, SAC, and FPA, respectively. This superiority stems from the conditional flow matching actor, which transports Gaussian priors over $K_{\rm FM}$ Euler steps to accurately capture complex multimodal high-reward action distributions in the coupled MA positioning and beamforming space. 

\begin{figure}[t]
  \centering
  \includegraphics[width=0.9\columnwidth]{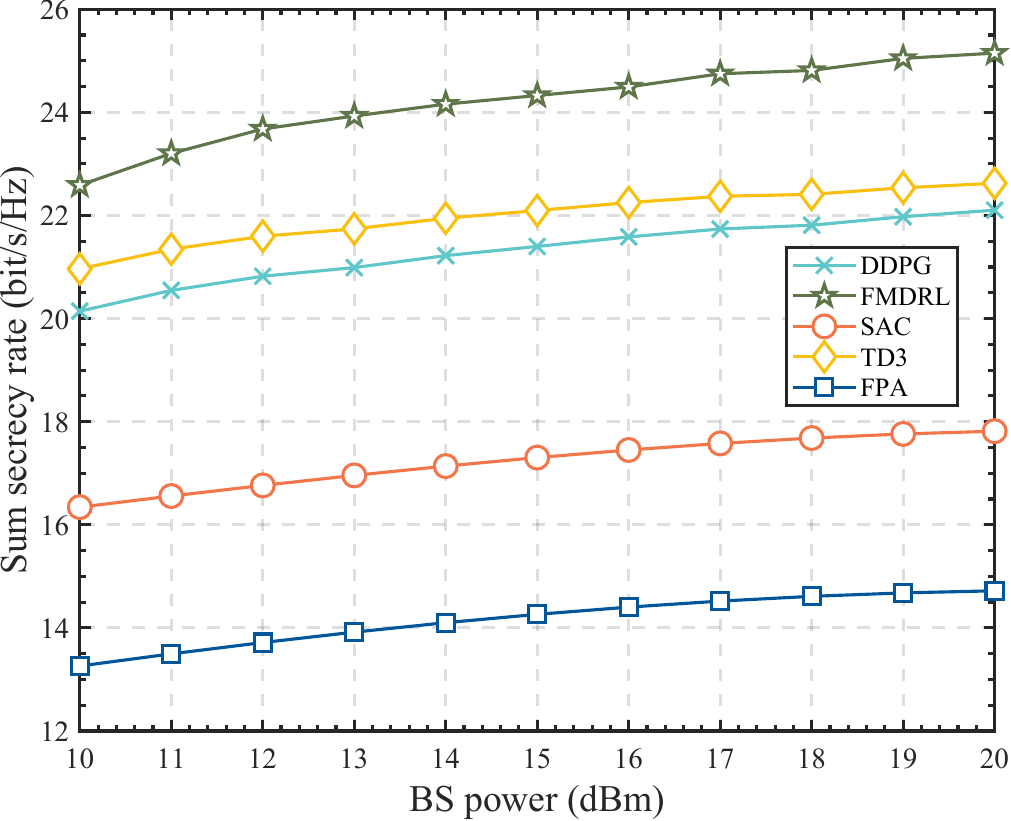}
 \caption{Sum secrecy rate versus the BS transmit power $P_{\rm BS}^{\max}$.}
  \label{fig:power_vs_rate}
\end{figure}

Fig.~\ref{fig:power_vs_rate} illustrates the sum secrecy rate versus the BS transmit power $P_{\rm BS}^{\max}$ ranging from $\text{10 dBm}$ to $\text{20 dBm}$. As observed in Fig.~\ref{fig:power_vs_rate}, the sum secrecy rates of all algorithms increase monotonically with the increase in $P_{\rm BS}^{\max}$, since a larger transmit power budget provides stronger downlink desired signals, greater artificial noise jamming capability, and higher radar echo SNR for EKF-based eavesdropper tracking. Across the entire power regime, FPA yields the lowest sum secrecy rate, whereas the MA algorithms exploit continuous spatial DoFs to reshape the steering vectors, thereby simultaneously enhancing legitimate channel gains and deepening spatial nulls toward the dynamic eavesdropper. Among all compared schemes, the proposed FMDRL consistently achieves the highest sum secrecy rate. Specifically, at $P_{\rm BS}^{\max}=\text{20 dBm}$, FMDRL reaches $\text{25.15 bit/s/Hz}$, achieving performance gains of $\text{11.18}\%$, $\text{13.80}\%$, $\text{41.13}\%$, and $\text{70.86}\%$ over TD3, DDPG, SAC, and FPA, respectively.

\begin{figure}[t]
  \centering
  \includegraphics[width=0.9\columnwidth]{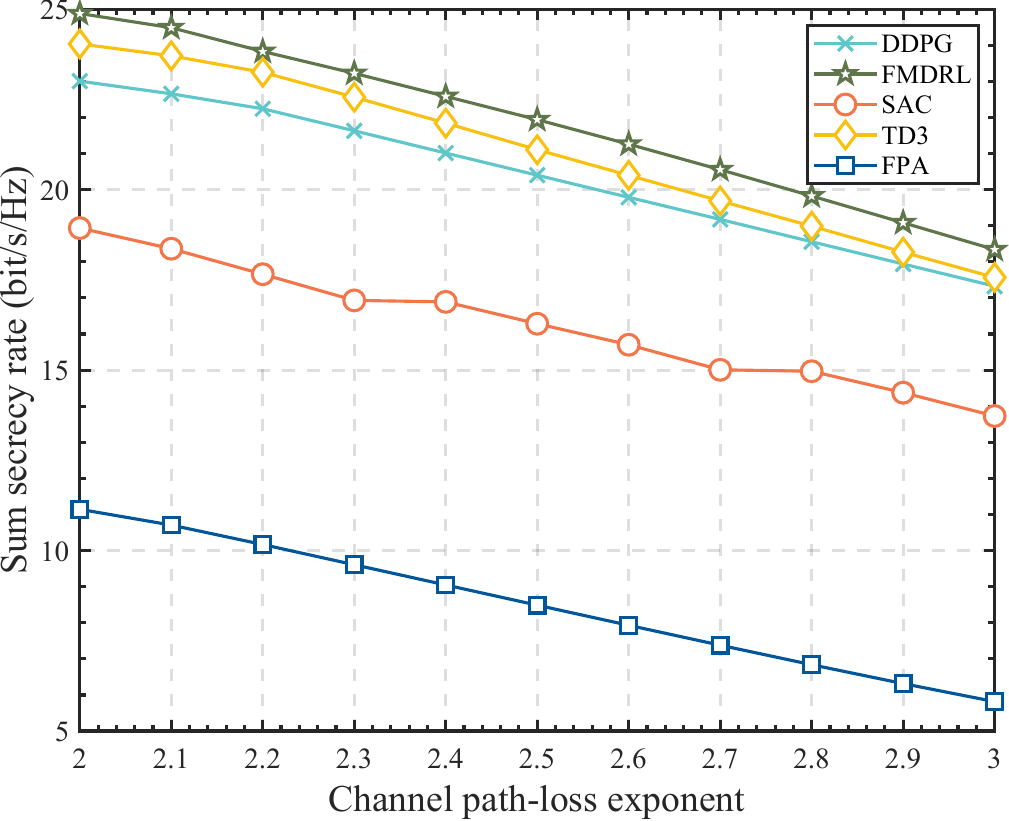}
\caption{Sum secrecy rate versus the channel path-loss exponent $o$.}
  \label{fig:path_loss_vs_rate}
\end{figure}

Fig.~\ref{fig:path_loss_vs_rate} depicts the sum secrecy rate versus the channel path-loss exponent $o$ varying from $\text{2.0}$ to $\text{3.0}$. As expected, the sum secrecy rates of all algorithms decrease monotonically with the increase in $o$, as a larger path-loss exponent induces more severe large scale signal attenuation over propagation links. Nevertheless, the proposed FMDRL consistently outperforms all baseline schemes across the entire range of $o$. It is worth noting that as $o$ increases from $\text{2.0}$ to $\text{3.0}$, FPA suffers a drastic relative performance loss of $\text{47.76}\%$, whereas FMDRL declines by only $\text{26.29}\%$. This phenomenon arises because under heavy path loss, both communication SINR and radar echo SNR become severely power-limited, which degrades EKF tracking precision unless compensated by spatial array gain. While FPA is confined to fixed array manifolds, the MA- algorithms dynamically reposition antenna elements to align with constructive multipath phases, and FMDRL further exploits its multimodal transport policy to identify globally superior antenna coordinates that compensate for severe path attenuation.

\begin{figure}[t]
  \centering
  \includegraphics[width=0.9\columnwidth]{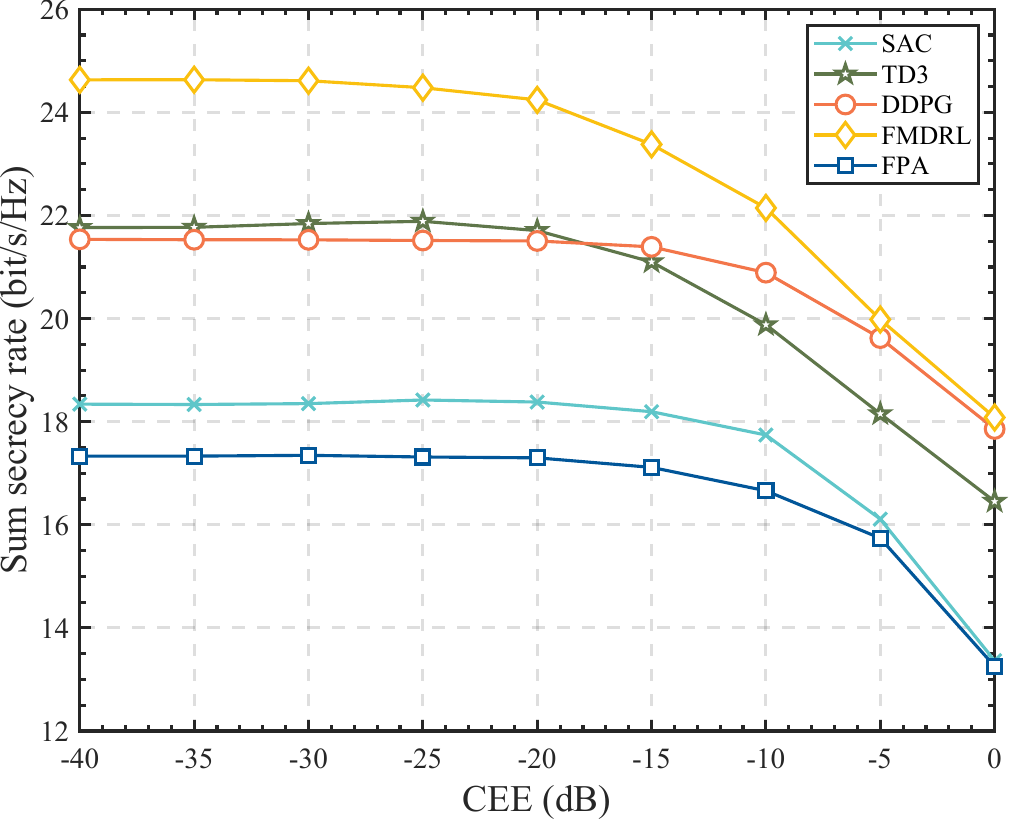}
\caption{Sum secrecy rate versus the CEE.}
  \label{fig:cee_vs_rate}
    \vspace{-1.5em}
\end{figure}

To evaluate the robustness of the proposed algorithms against imperfect CSI, Fig.~\ref{fig:cee_vs_rate} illustrates the sum secrecy rate versus the channel estimation error (CEE) ranging from $-\text{40 dB}$ to $\text{0 dB}$. The
CEE can be expresssed as follows
\begin{equation}
\text{CEE} = 10 \log_{10} \left( 
\frac{\mathbb{E}\left[\lVert \mathbf{h} - \hat{\mathbf{h}} \rVert_2^2 \right]}
{\mathbb{E}\left[\lVert \mathbf{h} \rVert_2^2 \right]}
\right),\label{CEE}
\end{equation}
where a lower CEE corresponds to higher channel estimation accuracy.
Under imperfect CSI, the estimation error vector $\mathbf{z}=\mathbf{h}-\hat{\mathbf{h}}$ follows a zero-mean circularly symmetric complex Gaussian distribution with spatially white covariance. A clear two stage trend can be observed across all curves: when the CEE lies within the low-error regime of $[-\text{40}, -\text{20}]\text{ dB}$, the sum secrecy rates of all algorithms remain nearly constant because minor channel perturbations are negligible relative to the spatial beamforming gain. However, once the CEE exceeds $-\text{20 dB}$, all curves exhibit an accelerated decline as severe CSI mismatch causes beam-pointing deviations, AN leakage into legitimate user channels, and incomplete self-interference cancellation. Throughout the entire CEE range, FMDRL achieves the highest sum secrecy rate. Moreover, two interesting baseline phenomena emerge in Fig.~\ref{fig:cee_vs_rate}. First, a performance crossover occurs between TD3 and DDPG: TD3 slightly outperforms DDPG when $\text{CEE}\leq-\text{20 dB}$, whereas DDPG surpasses TD3 when $\text{CEE}\geq-\text{15 dB}$. This is because TD3 injects target policy smoothing noise and adopts pessimistic twin-critic clipping. When large CEE already corrupts the state features, this additional target perturbation compounds with channel uncertainty and renders TD3 overly conservative. Second, as CEE approaches $\text{0 dB}$, SAC degrades nearly to the level of FPA, indicating that a high-entropy unimodal Gaussian policy fails to preserve precise spatial nulls under severe CSI errors, thereby losing most of the MA positioning gain.

\vspace{-1em}
\begin{figure}[H]
  \centering
  \includegraphics[width=0.9\columnwidth]{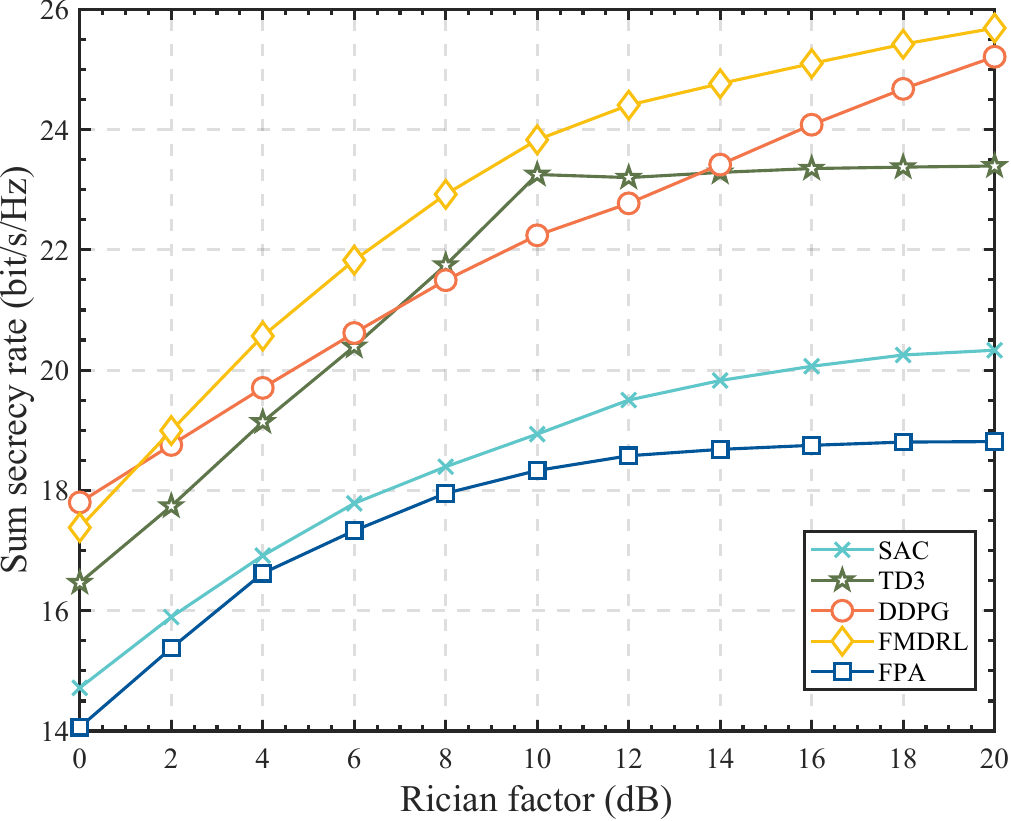}
\caption{Sum secrecy rate versus the Rician factor.}
  \label{fig:rician_vs_rate}
    \vspace{-1em}
\end{figure}

Finally, Fig.~\ref{fig:rician_vs_rate} presents the sum secrecy rate versus the Rician factor varying from $\text{0 dB}$ to $\text{20 dB}$. As shown in Fig.~\ref{fig:rician_vs_rate}, the sum secrecy rates of all algorithms improve with the increase in the Rician factor, since a stronger LoS component reduces random NLoS multipath scattering and renders the spatial channel structure more directional and predictable for joint dynamic beam tracking and MA positioning. At $\text{0 dB}$, where NLoS scattering is prominent, DDPG and FMDRL exhibit comparable performance. However, as the Rician factor increases from $\text{2 dB}$ to $\text{20 dB}$, FMDRL consistently surpasses all baseline algorithms. As the LoS component becomes dominant, the array response $\bm{\upalpha}(\vect P,\vect u)$ creates periodic, highly multimodal constructive and destructive interference patterns along the linear array aperture, which FMDRL can effectively navigate via its conditional flow matching velocity field. Notably, TD3 grows rapidly from $\text{0 dB}$ to $\text{10 dB}$, but saturates at approximately $\text{23.39 bit/s/Hz}$ when the Rician factor exceeds $\text{10 dB}$, eventually being overtaken by DDPG. This saturation occurs because in strongly LoS-dominated channels, small variations in MA positions induce sharp phase shifts in the steering vectors. The target policy smoothing noise in TD3 blurs these sharp geometric optima and traps the array in a local spatial mode. Similarly, FPA saturates above $\text{12 dB}$  because with fixed antenna spacing, reducing NLoS randomness alone cannot alter the intrinsic LoS steering vector correlation between legitimate users and the eavesdropper. 

\subsection{Sensing Performance}
In this subsection, the proposed algorithm FMDRL is compared with several baseline algorithms in terms of sensing metrics. In addition, the beam pattern results of the proposed algorithm at different time slots are analyzed. The experimental results show that the proposed algorithm achieves good sensing performance while taking into account communication performance.

\begin{figure}[t]
    \centering
    \subfloat[Azimuth tracking RMSE.\label{fig:degree_rmse}]{
        \includegraphics[width=0.9\columnwidth]{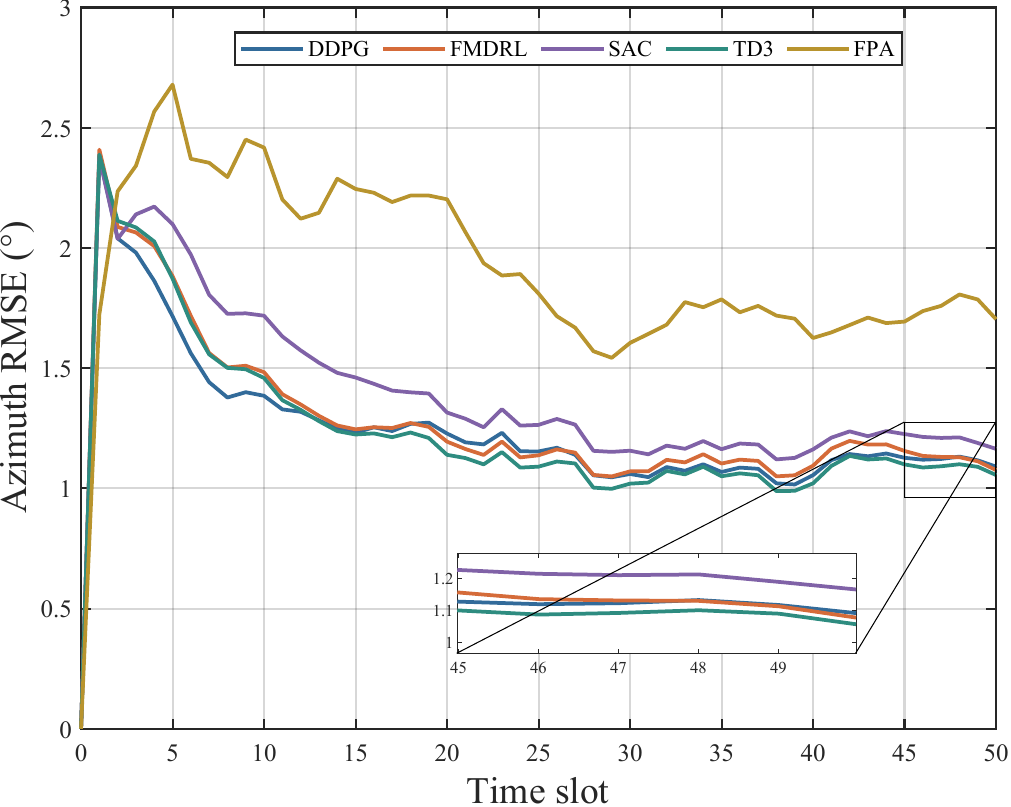}
    }

    \vspace{1mm}

    \subfloat[Distance tracking RMSE.\label{fig:distance_rmse}]{
        \includegraphics[width=0.9\columnwidth]{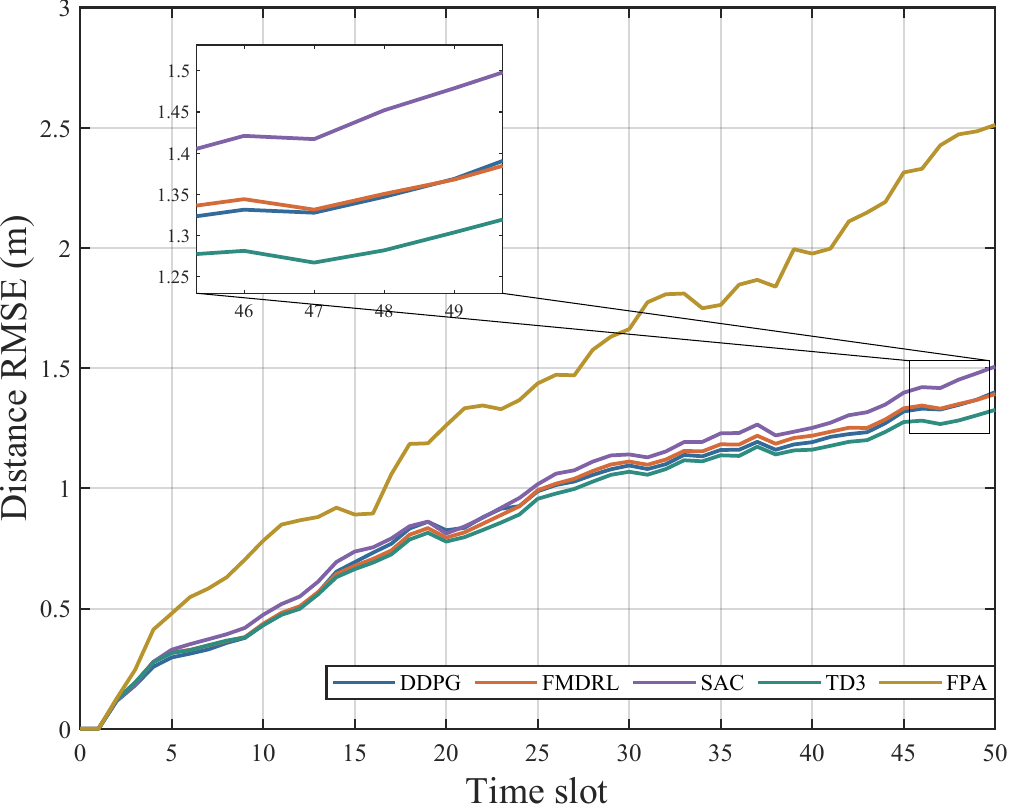}
    }
    \caption{Tracking RMSE versus time slots across different algorithms.}
    \label{fig:tracking_rmse}
    \vspace{-1.5em}
\end{figure}

Fig.~\ref{fig:tracking_rmse} illustrates the azimuth and distance tracking RMSEs across different algorithms. As shown in Fig.~\ref{fig:degree_rmse}, the azimuth RMSEs decrease rapidly during the initial slots and gradually stabilize as the EKF accumulates radar observations. In contrast, the distance RMSEs in Fig.~\ref{fig:distance_rmse} exhibit an upward trend, primarily due to the increasing two-way path loss as the eavesdropper moves away from the BS, which weakens radar echoes and degrades delay estimation accuracy. FPA consistently exhibits the poorest tracking performance, highlighting the limitations of its fixed array geometry. By dynamically adjusting antenna positions, MA algorithms enhance spatial resolution and radar echo gain, thereby improving tracking accuracy. Specifically, FMDRL reduces the average azimuth RMSE by $\text{33.71}\%$ and the terminal distance RMSE by $\text{44.59}\%$ compared with FPA. Notably, TD3 and DDPG achieve marginally better angular tracking accuracy, while TD3 also attains a slightly lower terminal distance RMSE. This observation reflects the inherent communication--sensing trade-off in Problem~\eqref{prob:P1}, where sensing accuracy is imposed as a constraint rather than directly optimized. Further improving tracking accuracy requires additional sensing resources, potentially increasing information leakage. Consequently, FMDRL maintains competitive sensing accuracy while prioritizing the maximization of the sum secrecy rate.

\begin{figure}[t]
  \centering
  \includegraphics[width=0.85\columnwidth]{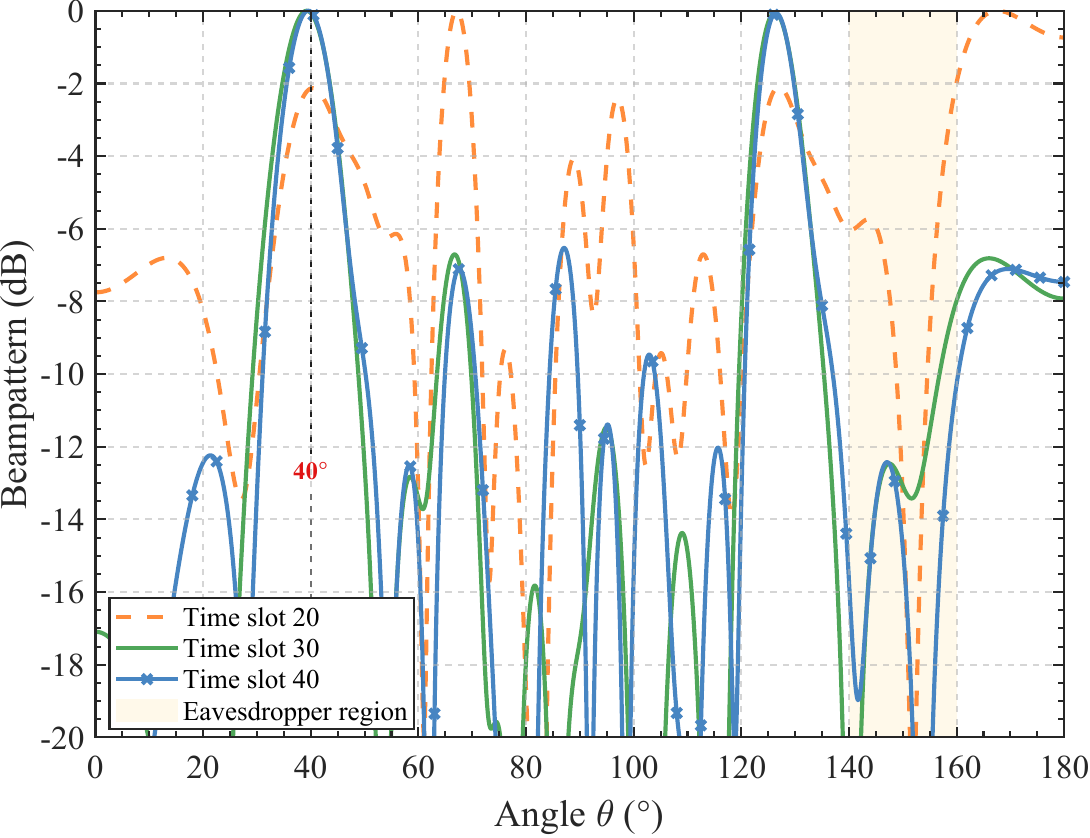}
  \caption{Transmit beampatterns of the proposed FMDRL algorithm at time slots $\text{20}$, $\text{30}$, and $\text{40}$.}
  \label{fig:tx_beampattern}
\end{figure}

To further reveal the spatial security and tracking mechanism of the proposed FMDRL algorithm, Fig.~\ref{fig:tx_beampattern} illustrates the normalized transmit beampatterns at time slots $\text{20}$, $\text{30}$, and $\text{40}$ across the azimuth range $\theta\in[\text{0}^{\circ},\text{180}^{\circ}]$. In Fig.~\ref{fig:tx_beampattern}, the vertical dashed line at $\theta=\text{40}^{\circ}$ marks the stationary downlink legitimate user direction, while the yellow shaded interval $\theta\in[\text{140}^{\circ},\text{160}^{\circ}]$ indicates the angular region traversed by the mobile eavesdropper during the late tracking stages. At time slot $\text{20}$, the radiation pattern exhibits relatively dispersed energy distribution, with a less pronounced main lobe toward the legitimate user at $\text{40}^\circ$ and noticeable radiation leakage within the eavesdropper region. As time progresses, FMDRL dynamically adjusts the transmit MA positions $\vect P_{\rm tx}[t]$ and beamformers $\vect W[t]$ to reshape the spatial radiation pattern. Specifically, the main lobe toward the legitimate user at $\text{40}^\circ$ becomes more concentrated and remains well aligned at subsequent time slots, ensuring effective signal transmission. Meanwhile, increasingly pronounced spatial nulls emerge within the eavesdropper region, with their locations adaptively adjusted to suppress radiation toward the moving eavesdropper. These observations demonstrate that FMDRL effectively exploits MA mobility and adaptive beamforming to maintain directional signal enhancement for legitimate users while mitigating information leakage toward the dynamic eavesdropper.

\begin{figure}[t]
  \centering
  \includegraphics[width=0.85\columnwidth]{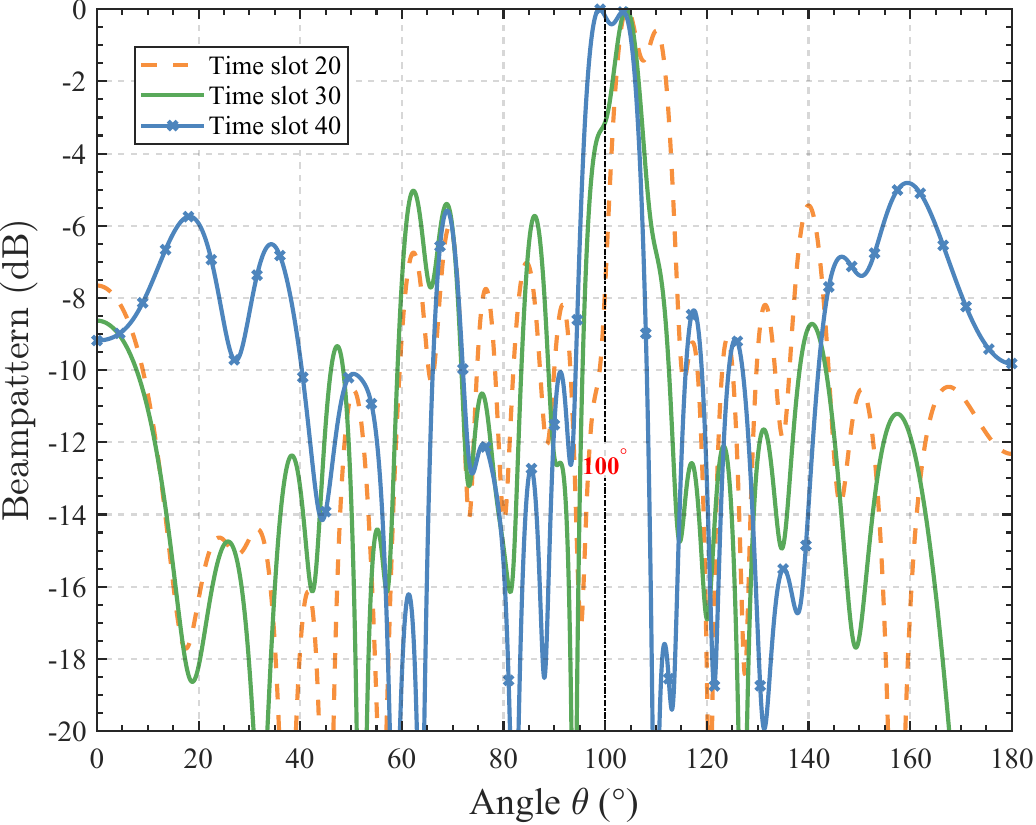}
  \caption{Receive beampatterns of the proposed FMDRL algorithm at time slots $\text{20}$, $\text{30}$, and $\text{40}$.}
  \label{fig:rx_beampattern}
  \vspace{-1.5em}
\end{figure}

Finally, Fig.~\ref{fig:rx_beampattern} presents the normalized receive beampatterns of the proposed FMDRL algorithm at time slots $\text{20}$, $\text{30}$, and $\text{40}$, where the vertical dashed line at $\theta=\text{100}^{\circ}$ denotes the azimuth of the legitimate uplink user. In the considered full-duplex ISAC system, the receive array must simultaneously capture the legitimate uplink signal and the target radar echo while suppressing strong residual self-interference coupled from the moving transmit MA. As observed in Fig.~\ref{fig:rx_beampattern}, despite the time-varying self-interference channel and target angle across slots $\text{20}$, $\text{30}$, and $\text{40}$, FMDRL continuously coordinates the receive MA positions $\vect P_{\rm rx}[t]$ and beamforming vector $\vect F[t]$ such that the primary receive main lobe remains tightly directed toward the legitimate uplink user at $\theta=\text{100}^{\circ}$, converging from $\text{104.3}^{\circ}$ at slot $\text{20}$ and $\text{104.1}^{\circ}$ at slot $\text{30}$ to $\text{98.9}^{\circ}\text{--}\text{100}^{\circ}$ with $\text{0 dB}$ normalized peak gain at slot $\text{40}$. Meanwhile, the receive sidelobes across non-user angular sectors are deeply suppressed below $-\text{10 dB}$ to $-\text{20 dB}$, accompanied. These results demonstrate that by jointly optimizing the receive antenna positioning and beamforming, FMDRL ensures high gain coherent uplink reception while effectively mitigating full-duplex residual SI and multipath clutter.

\subsection{Parameter Sensitivity Analysis}
In this subsection, we investigate the sensitivity of the proposed FMDRL algorithm to the number of ODE solver steps $K_{\rm FM}$, focusing on the continuous flow evolution of optimization variables and the resulting communication and sensing behaviors.

\begin{figure}[t]
  \centering
  \includegraphics[width=0.85\columnwidth]{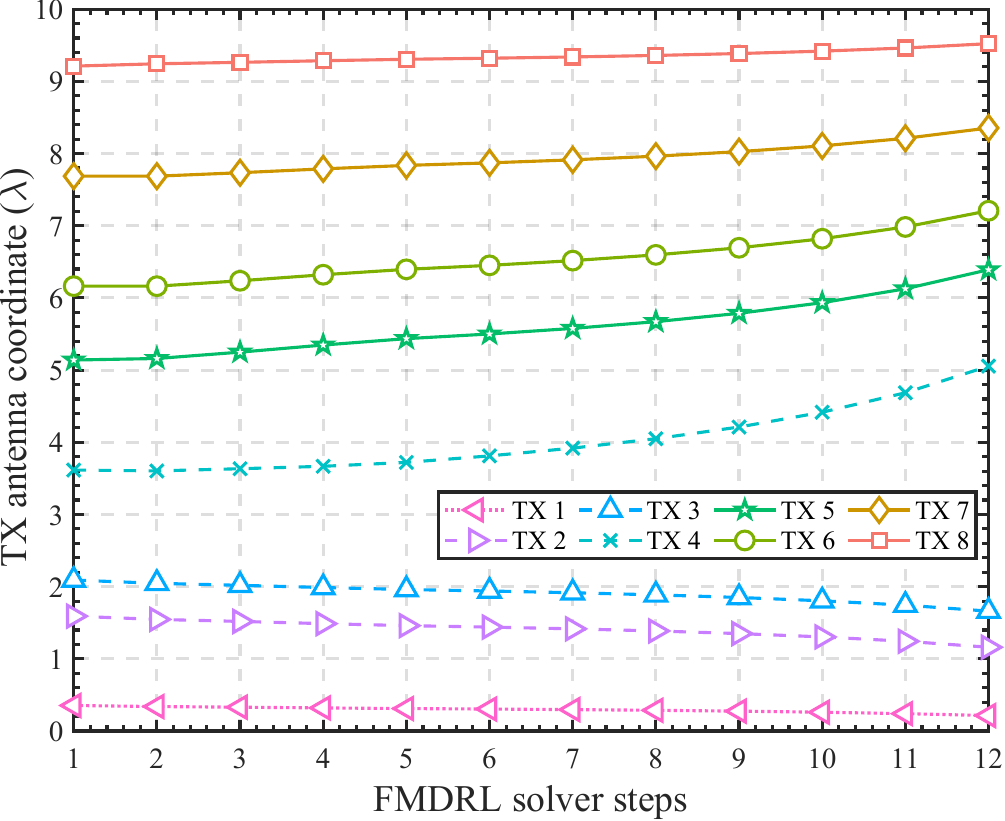}
  \caption{Evolution of transmit MA coordinates across FMDRL solver steps.}
  \label{fig:coordinate_vs_step}
\end{figure}

We first examine the internal action generation process of the conditional flow matching actor across integration steps. Fig.~\ref{fig:coordinate_vs_step} illustrates the trajectories of the eight transmit MA coordinates from solver step $\text{1}$ to step $\text{12}$. As governed by the explicit Euler recursion in~\eqref{eq:fmdrl_euler} and the order-preserving projection decoder in~\eqref{eq:fmdrl_position_mapping}, the state conditioned velocity network $\vect v_{\bm{\uptheta}}(\vect x^{(j)},\tau_j,\vect s)$ progressively transports the initial reference iterate toward the high value action manifold. As observed in Fig.~\ref{fig:coordinate_vs_step}, all eight antenna trajectories evolve smoothly and monotonically across the solver steps without any trajectory crossing, strictly preserving the minimum spacing $d_{\min}$ throughout the flow transport. Specifically, as the flow step progresses, the intermediate elements such as TX~4 and TX~5 exhibit more pronounced spatial adjustments in the later integration steps. This smooth divergence without trajectory intersections visually demonstrates how the learned velocity field progressively refines a coarse initial placement into an expanded array geometry with precise antenna positioning, thereby enhancing system performance.
\begin{figure}[t]
  \centering
  \includegraphics[width=0.85\columnwidth]{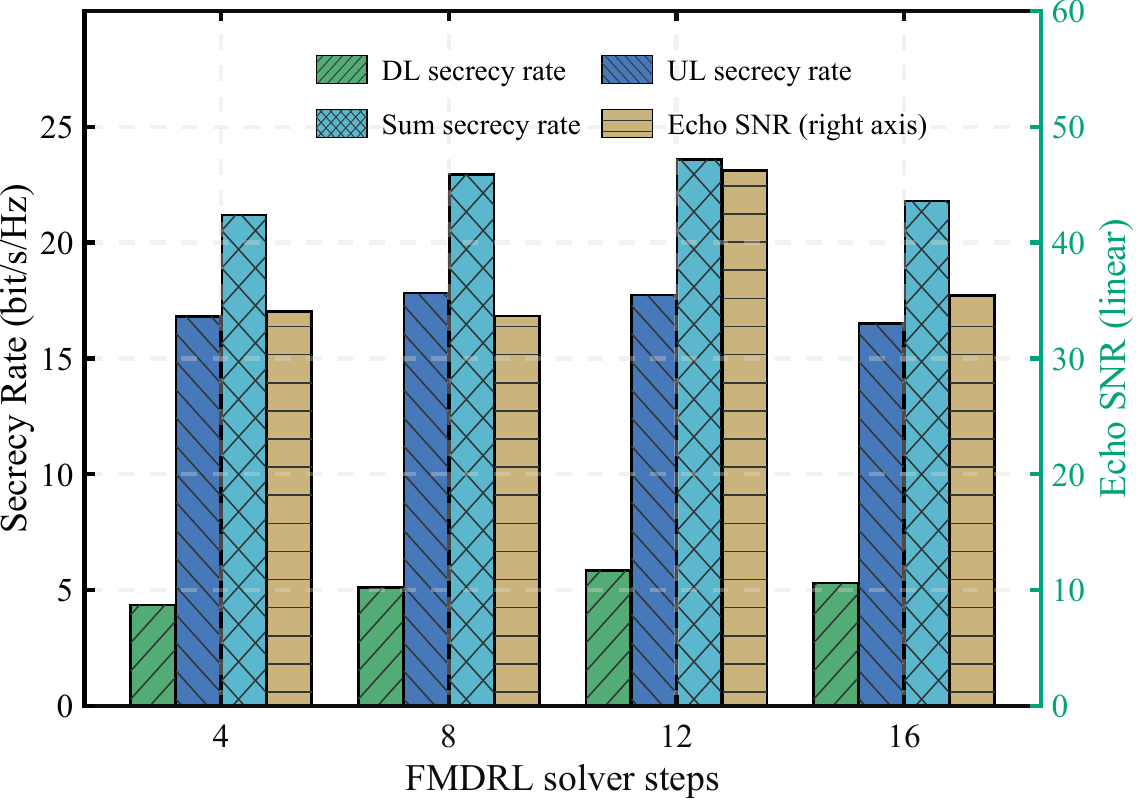}
  \caption{Comparison of secrecy rates and radar echo SNR under different numbers of FMDRL solver steps.}
  \label{fig:solve_step_bar}
  \vspace{-1.5em}
\end{figure}

Fig.~\ref{fig:solve_step_bar} compares the downlink secrecy rate, uplink secrecy rate, sum secrecy rate, and radar echo SNR under different numbers of solver steps $K_{\rm FM}\in\{\text{4},\text{8},\text{12},\text{16}\}$. As observed, all four metrics improve steadily as $K_{\rm FM}$ increases from $\text{4}$ to $\text{12}$, before experiencing a slight decline at $K_{\rm FM}=\text{16}$. When $K_{\rm FM}$ is small, the coarse Euler discretization limits the fidelity of continuous action generation. Increasing $K_{\rm FM}$ refines the numerical integration of the learned velocity field, thereby facilitating more precise joint adjustment of MA positions, transmit and receive beamforming, artificial noise, and uplink transmit power. The results indicate that $K_{\rm FM}=\text{12}$ provides sufficient integration resolution for the considered system, beyond which finer discretization yields diminishing returns. The slight decline at $K_{\rm FM}=\text{16}$ occurs because deeper unrolled Euler chains do not improve the underlying velocity field or critic approximations, but instead alter backpropagation dynamics and accumulate minor estimation errors, which perturb the tightly coupled beamforming, self-interference suppression, and echo tracking actions. Consequently, $K_{\rm FM}=\text{12}$ achieves the most favorable trade-off between decision quality and computational overhead, requiring fewer velocity-network evaluations than $K_{\rm FM}=\text{16}$.

\begin{figure}[t]
  \centering
  \includegraphics[width=0.85\columnwidth]{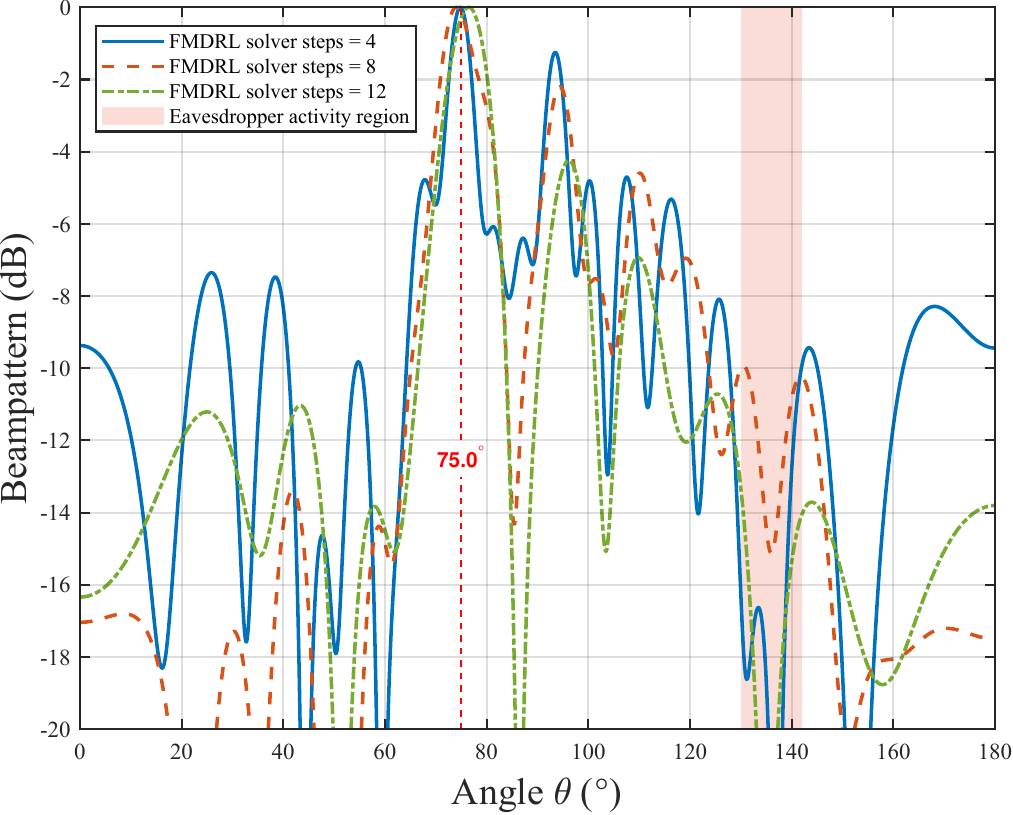}
  \caption{Transmit beampatterns of FMDRL under different numbers of FMDRL solver steps.}
  \label{fig:beampattern_vs_step}
  \vspace{-1.5em}
\end{figure}

Finally, Fig.~\ref{fig:beampattern_vs_step} depicts the normalized transmit beampatterns under solver steps $K_{\rm FM}\in\{\text{4},\text{8},\text{12}\}$, where the vertical dashed line at $\theta=\text{75.0}^{\circ}$ marks the legitimate user direction and the shaded band indicates the eavesdropper activity region. A clear progressive refinement in both the main lobe and sidelobes can be observed as $K_{\rm FM}$ increases. When $K_{\rm FM}=\text{4}$, coarse discretization inaccuracies distort the main lobe near $\theta=\text{75.0}^{\circ}$ and induce pronounced sidelobe fluctuations. As $K_{\rm FM}$ increases to $\text{8}$ and $\text{12}$, the multi step velocity field iteratively corrects the coupled antenna positions and complex beamforming phases along the flow trajectory. As a result, the main lobe becomes increasingly symmetric and concentrically aligned with the desired direction $\theta=\text{75.0}^{\circ}$. Simultaneously, the overall sidelobe envelope across non-user sectors is substantially lowered and smoothed, while maintaining consistent radiation suppression within the eavesdropper activity region. This confirms that sufficient steps are essential for translating the velocity field updates into precise sensing performance.

\section{Conclusion}\label{sec:conclusion}
This paper has investigated a joint antenna positioning and dynamic beam tracking algorithm for the MA-enabled full-duplex secure ISAC system in the presence of a dynamic eavesdropper. First, we established an MA secure ISAC framework and employed an EKF framework to track the mobile eavesdropper via radar echoes. By comprehensively considering pysical constraints, a sequential optimization problem was formulated with the objective of maximizing the system sum secrecy rate. Subsequently, we proposed the FMDRL optimization algorithm incorporated with deterministic action decoding and penalty-based constraint handling strategies. By integrating a conditional flow matching velocity network with twin critics, this algorithmic framework was applied to the joint design of optimization variables. Finally, extensive numerical simulations were conducted to compare the proposed algorithm against several DRL and FPA baseline. The experimental results verified the rationality of the flow matching action generation mechanism and the effectiveness of the proposed algorithm in balancing the trade-off between communication security and target tracking accuracy, offering new insights for dynamic MA-enabled secure ISAC systems.

\balance

\bibliographystyle{IEEEtran}
\bibliography{references_intro}

\end{document}